\documentclass [11pt]{article}
\usepackage{jheppub}
\usepackage{amsfonts}
\usepackage{graphicx}
\usepackage{amsmath,amssymb, amsfonts,color}
\usepackage{latexsym}
\usepackage{mathrsfs}
\usepackage{bbold}
\usepackage{tabularx}
\usepackage{color}
\usepackage{slashed}
\usepackage{cancel}
\usepackage{soul} %
\usepackage{setspace} %
\usepackage{comment}
\usepackage{booktabs}
\usepackage{multirow}
\usepackage{tikz-feynman}
\tikzfeynmanset{compat=1.0.0}
\usepackage{epsfig}
\usepackage{slashed}
\usepackage{epstopdf}
\usepackage{latexsym}
\usepackage{subcaption}
\usepackage{mdframed}
\usepackage{bm}
\usepackage{upgreek}

\usepackage{tikz}
\usetikzlibrary{decorations.pathmorphing}
\pgfdeclaredecoration{complete sines}{initial}
{
    \state{initial}[
        width=+0pt,
        next state=sine,
        persistent precomputation={\pgfmathsetmacro\matchinglength{
            \pgfdecoratedinputsegmentlength / int(\pgfdecoratedinputsegmentlength/\pgfdecorationsegmentlength)}
            \setlength{\pgfdecorationsegmentlength}{\matchinglength pt}
        }] {}
    \state{sine}[width=\pgfdecorationsegmentlength]{
        \pgfpathsine{\pgfpoint{0.25\pgfdecorationsegmentlength}{0.5\pgfdecorationsegmentamplitude}}
        \pgfpathcosine{\pgfpoint{0.25\pgfdecorationsegmentlength}{-0.5\pgfdecorationsegmentamplitude}}
        \pgfpathsine{\pgfpoint{0.25\pgfdecorationsegmentlength}{-0.5\pgfdecorationsegmentamplitude}}
        \pgfpathcosine{\pgfpoint{0.25\pgfdecorationsegmentlength}{0.5\pgfdecorationsegmentamplitude}}
}
    \state{final}{}
}

\usepackage{rotating}
\usepackage{bbm}
\usepackage{bbold}
\usepackage{enumitem}

\usepackage{empheq}
\usepackage{caption}

\usepackage{amsthm}

\theoremstyle{definition}

\newtheorem {remark} {Remark}[section]

\definecolor{myblue}{rgb}{.8, .8, 1}

\definecolor{red}{rgb}{1,0,0}
\usetikzlibrary{positioning}
\tikzset{
    mybluenode/.style={
        draw=black, circle, minimum width=2cm, inner sep=0pt
        },
    myblacknode/.style={
        circle, inner sep=1pt, fill=black
        },
    }

\newcommand{\ra}[1]{\renewcommand{\arraystretch}{#1}}

\newcommand*\be{\beta}

\newcommand*\ka{\kappa}

\newcommand{\bea}{\begin{equation} \begin{aligned}} \newcommand{\eea}{\end{aligned} \end{equation}}
\def\be{\begin{equation}} \def\ee{\end{equation}} 
\def\beq{\begin{equation}} \def\eeq{\end{equation}}

\newcommand{\eps}{\epsilon}

\usepackage[english]{babel}
\usepackage{csquotes}
\MakeOuterQuote{"}

\makeatletter
\def\@fpheader{\ }
\makeatother

\title{Disturbing news about the $d=2+\epsilon$ expansion}
\author{Fabiana De Cesare,}
\author{Slava Rychkov}

\affiliation{Institut des Hautes \'Etudes Scientifiques, 91440 Bures-sur-Yvette, France}
\emailAdd{decesare@ihes.fr}
\emailAdd{slava@ihes.fr}
	
\abstract{The $O(N)$ Non-Linear Sigma Model (NLSM) in $d=2+\epsilon$ has long been conjectured to describe the same conformal field theory (CFT) as the Wilson-Fisher (WF) $O(N)$ fixed point obtained from the $\lambda(\phi^2)^2$ model in
$d=4-\epsilon$. In this work, we put this conjecture into question, building on the recent observation~\cite{Jones:2024ept} that the NLSM CFT possesses a protected operator with dimension $N-1$, which is instead absent in the WF $O(N)$ CFT. We investigate the possibility of lifting this operator via multiplet recombination — the only known mechanism that could resolve this mismatch while preserving a connection between the two theories. We also explore an alternative scenario in which the NLSM $O(N)$ fixed point in $d=2+\epsilon$ is not continuously connected to the WF $O(N)$ CFT, and instead corresponds to a different universality class. For $N=3$, this could be related to the hedgehog-suppressed critical point, which describes the Néel–VBS phase transition in 3D.}

\begin{document}
	\maketitle
	\flushbottom
	
\section{Introduction}

One of the most famous universality classes of 3D phase transitions is the $O(N)$ universality class. Its critical exponents can be accessed by two famous expansions from $4-\eps$ and from $2+\eps$ dimensions. Or can they? Before we explain what we mean by this provocative question, let us review some basic facts and definitions.
Two well-known representatives of the 3D $O(N)$ universality class are:
\begin{itemize}
	\item
 The $O(N)$ vector model, which is a lattice model described by the Hamiltonian:
\beq
\label{eq:HON}
H = -\beta \sum\nolimits_{\langle xy\rangle} \vec n_x\cdot \vec n_y\,\qquad |\vec n_x|=1,
\eeq
where $\vec n_x\in S^{N-1}$ are $N$-component unit-length vectors at the vertices of the cubic lattice $x\in \mathbb{Z}^3$, with nearest-neighbor ferromagnetic interactions.
\item
The field theory of an $N$-component scalar field $\vec \phi(x)\in \mathbb{R}^N$ with $O(N)$ invariant quartic interactions:
\beq
\label{eq:SWF}
S = \int d^3 x\ (\partial_\mu \vec\phi)^2 + m^2 \vec \phi^2+\lambda (\vec \phi^2)^2\,.
\eeq
\end{itemize}
That these models have a second-order phase belonging to the same universality class is supported by a large body of numerical and theoretical considerations. From a modern perspective, this universality class is described by a unitary 3D CFT called here the "3D Bootstrap $O(N)$ CFT", to stress that it can be studied using the conformal bootstrap \cite{Poland:2018epd,Kos:2015mba,Chester:2019ifh,Chester:2020iyt}, directly in 3D. We will use "3D Bootstrap $O(N)$ CFT" and "3D $O(N)$ universality class" as synonyms. Table \ref{tab:theories} summarizes this and subsequent terminology. We consider $N$ to be an integer for most of the paper, except for some comments in Section \ref{sec:comments-on-non-integer-n}.

\begin{table}
	\centering
	\begin{tabular}{ll}
		\toprule
		Theory & Definition \\\midrule
		(1)\ 3D $O(N)$ universality class & Phase transition of \eqref{eq:HON} and \eqref{eq:SWF}\\
		(2)\ 3D Bootstrap $O(N)$ CFT & Synonymous to (1)\\
		(3)\ WF $O(N)$ CFT, $d<4$ & Analytic continuation of \eqref{eq:SWF} from $d=4-\epsilon$ \\
		(4)\ NSLM $O(N)$ CFT, $d>2$ & Analytic continuation of NLSM \eqref{eq:action0} from $d=2+\epsilon$\\
		(5) Large-$N$ $O(N)$ CFT, $2<d<4$ & Large-$N$ expansion
		 \\\bottomrule
	\end{tabular}
	\caption{Various theories which will occur in our discussion. Since the scalar $\phi^4$ theory is UV complete, (3) has to coincide in 3D with (1). We would like to examine whether this also holds for (4). Theory (5) will be brought up in Section \ref{sec:presentation}.} 
	\label{tab:theories}
\end{table}

Before the conformal bootstrap came to the fore, these models have been studied using different techniques. One time-honored idea is analytic continuation in dimension. Wilson and Fisher  \cite{Wilson:1971dc,Wilson:1973jj} famously analytically continued field theory \eqref{eq:SWF} to non-integer $d$ and showed that for $d=4-\epsilon$ the phase transition can be studied in perturbation theory. More generally, this procedure gives rise to a family of conformal fixed points in $2<d<4$ dimensions, called here "WF $O(N)$ CFT". Critical exponents can be computed in a power series expansion in $\epsilon$, then continued to $\epsilon=1$, in excellent agreement with other numerical techniques working directly in 3D, like lattice Monte Carlo simulations and the bootstrap.

On the other hand, Polyakov \cite{Polyakov:1975rr} considered the Non-Linear Sigma Model (NLSM), which is the field theory with the action obtained as the naive continuum limit of \eqref{eq:HON}:
\begin{equation}
	S=\frac{1}{2t} \int d^{d}x (\partial_\mu  \vec n(x))^2\,,\quad |\vec n(x)|=1\,.
	\label{eq:action0}
\end{equation}
For $d=2$, the model is renormalizable. Polyakov computed its leading beta function and found that the model is asymptotically free for $N>2$, due to positive curvature of the sphere \cite{Polyakov:1975rr}. 
From now on we will focus on the range $N>2$. Going to $d>2$, Polyakov realized that the NLSM should have a UV fixed point, which becomes weakly coupled in $d=2+\eps$ dimensions, see App.~\ref{app:review} for a review. He conjectured that for $\eps=1$ this fixed point should belong to the $O(N)$ universality class. Note that in 3D field theory \eqref{eq:action0} is non-renormalizable, so its precise definition requires a cutoff. With a UV cutoff, \eqref{eq:action0} is basically equivalent to \eqref {eq:HON}, so viewed this way Polyakov's conjecture would be non-controversial although not very useful computationally. That is however not how it is usually viewed. The nontrivial interpretation of Polyakov's conjecture is that the critical exponents can now be computed in a series expansion in $d=2+\eps$ and then continued to $\eps=1$. Such computations were performed by Br\'ezin, Zinn-Justin and others  \cite{Brezin:1975sq,Bardeen:1976zh,Brezin:1976ap,Brezin:1976qa,Hikami:1977vr}. They thus define a family of CFTs in $d>2$, called here "NLSM $O(N)$ CFT". {\bf Whether this $2+\epsilon$ expansion is supposed to give in $d=3$ the theory in the 3D $O(N)$ universality class is not fully obvious.} In fact this is a central question that we would like to scrutinize in this paper.

 Note that the $2+\epsilon$ expansion is a standard technique, well covered in textbooks and reviews \cite{ZJ, Cardy-book,amit2005field,Henriksson:2022rnm}. In this work we do not question the assumption that the technique defines a family of $d$-dimensional CFTs.\footnote{ We also accept the usual assumption of having full conformal invariance, as opposed to only scale invariance. While examples of physical theories with scale without conformal invariance are known \cite{Riva:2005gd,Mauri:2021ili,Gimenez-Grau:2023lpz}, it is understood that this behavior is non-generic as it requires the existence of a virial current \cite{Polchinski:1987dy}; see \cite{Nakayama:2013is} for a review.} Nor do we try to modify the technique in any way. Our conclusions will be based on following the technique to the letter, and performing some consistency checks against other known results.

Over the course of the years, other people sometimes wondered about the status of the $2+\eps$ expansion. At a purely numerical level, it has not matched the level of success of the $4-\epsilon$ expansion in reproducing the 3D critical exponents. More theoretically, in $d = 2 + \epsilon$ there exists a family of conformal primaries with large negative anomalous dimensions, at least at leading orders in $\epsilon$ \cite{Wegner:1987gu,Castilla_1997,Brezin:1996ff}. This raises concerns about the RG stability of this fixed point. Indeed the 3D $O(N)$ universality class has only one relevant singlet scalar operator. These worries could be dismissed by saying that the $2+\epsilon$ expansion is just more poorly summable than the $4-\epsilon$ expansion, and that these inconsistencies would perhaps go away as more orders were included. Unfortunately, the number of available perturbative orders in $d=2+\eps$ has stagnated, while there was continuing progress on the $4-\eps$ side.

More recently, another concern was raised in Ref.~\cite{Nahum_2015} for $N=3$, which noted that the $2+\epsilon$ expansion presumably cannot account for the topological defects (hedgehogs) which would be present in the 3D lattice formulation of the $O(3)$ vector model,\footnote{Also known as the Heisenberg model.} and should therefore connect to an $O(3)$-symmetric phase transition where the hedgehog configurations are suppressed, which belongs to a different universality class, described by the NCCP$^1$ model \cite{PhysRevLett.71.1911,Motrunich:2003fz}. This was further sharpened in \cite{Jones:2024ept} who realized that the NLSM $O(3)$ CFT in $d = 2 + \epsilon$ possesses a protected antisymmetric two-index tensor operator $B_{\mu\nu}$ with scaling dimension exactly equal to 2. When continued to 3D, this operator can be Hodge-dualized and becomes an additional conserved $U(1)$ current.
This lends support to the scenario of \cite{Nahum_2015}. Indeed, the 3D NCCP$^1$ model has an additional $U(1)$ current, coupled to the topological charge, while the 3D $O(3)$ universality class has no such currents. A similar concern arises for larger values of $N$, as the NLSM $O(N)$ CFT in 
$2+\epsilon$ still possesses a protected operator, in this case with dimension 
$N-1$, that is absent in the $4-\epsilon$ expansion. 

Given the accumulation of concerns,\footnote{ For completeness, we also mention the work of Cardy and Hamber \cite{Cardy-Hamber} who argued that vortices which play a role in the BKT transition for $N=d=2$ should also be accounted for $N>2$ and $d>2$. They proposed a beta-function in $d=2+\eps$ with extra couplings for vortex perturbations. This beta-function has never been derived from first principles. Vortex operators, being pointlike defects in $d=2$, should perhaps become dimension $\eps$ defects in $d=2+\eps$. In the latter case they are no longer local operators and the Cardy-Hamber proposal loses ground. Anyhow, their proposal amounts to modifying the $2+\eps$ expansion. As mentioned, in this work we want instead to examine its standard textbook version.} the time is ripe for a comprehensive discussion. It appears plausible that the WF $O(N)$ CFT depends analytically on $d$ at least in the range $3\le d<4$. Furthermore in 3D the first three theories in Table \ref{tab:theories} all agree:
\beq
\text{3D $O(N)$ universality class\  =\  3D Bootstrap $O(N)$ CFT = WF $O(N)$ CFT in $d=3$.}
\eeq 
But how does the NLSM $O(N)$ CFT fit into this picture? We will explore a few scenarios:
\begin{enumerate}
    \item \textbf{Analytic Connection:} The $2+\eps$ and $4-\eps$ expansions are analytically connected, so that
    \beq
    \text{NLSM $O(N)$ CFT $=$ WF $O(N)$ CFT for all $2<d<4$}\quad (?)
    \eeq 
    This used to be the most plausible scenario, but as mentioned above it is inconsistent with the presence of the extra protected operator in the NSLM $O(N)$ CFT, which is instead absent in the WF $O(N)$ CFT.
    \item \textbf{3D Intersection:} The two theories are distinct CFTs across the interval of $d$ but coincide specifically at $d=3$. This scenario is in principle possible for $N > 4$, where the protected operator is evanescent and disappears in 3D.
    \item\textbf{Continuous Connection:} The protected operator in the NLSM $O(N)$ CFT is lifted through multiplet recombination, enabling a continuous but non-analytic connection to the WF $O(N)$ CFT. To get rid of the protected operator in 3D, recombination should happen in $2<d<3$ for $N=3,4$, while it may happen for any $2<d<4$ for $N>4$.
    
    \item \textbf{Distinct Theories:} The NLSM and WF $O(N)$ CFTs follow entirely separate paths in the theory space and remain unrelated throughout the interval  $2<d<4$. This is in principle allowed for all finite values of $N$.
\end{enumerate}
The paper is structured as follows. In Section \ref{sec:presentation} we discuss the first two scenarios. In Section \ref{sec:recomb} we explore the third scenario, by considering
the lightest operators potentially involved in recombination. Finally, in Section \ref{sec:assessment}, we explore the last scenario and assess the situation, across different values of $N$. We conclude by outlining potential future directions, focusing especially on the case $N=3$.  

\section{Presentation of the problem and possible scenarios}
\label{sec:presentation}
\begin{figure}[h]  
	\centering
	\includegraphics[width=0.6\textwidth]{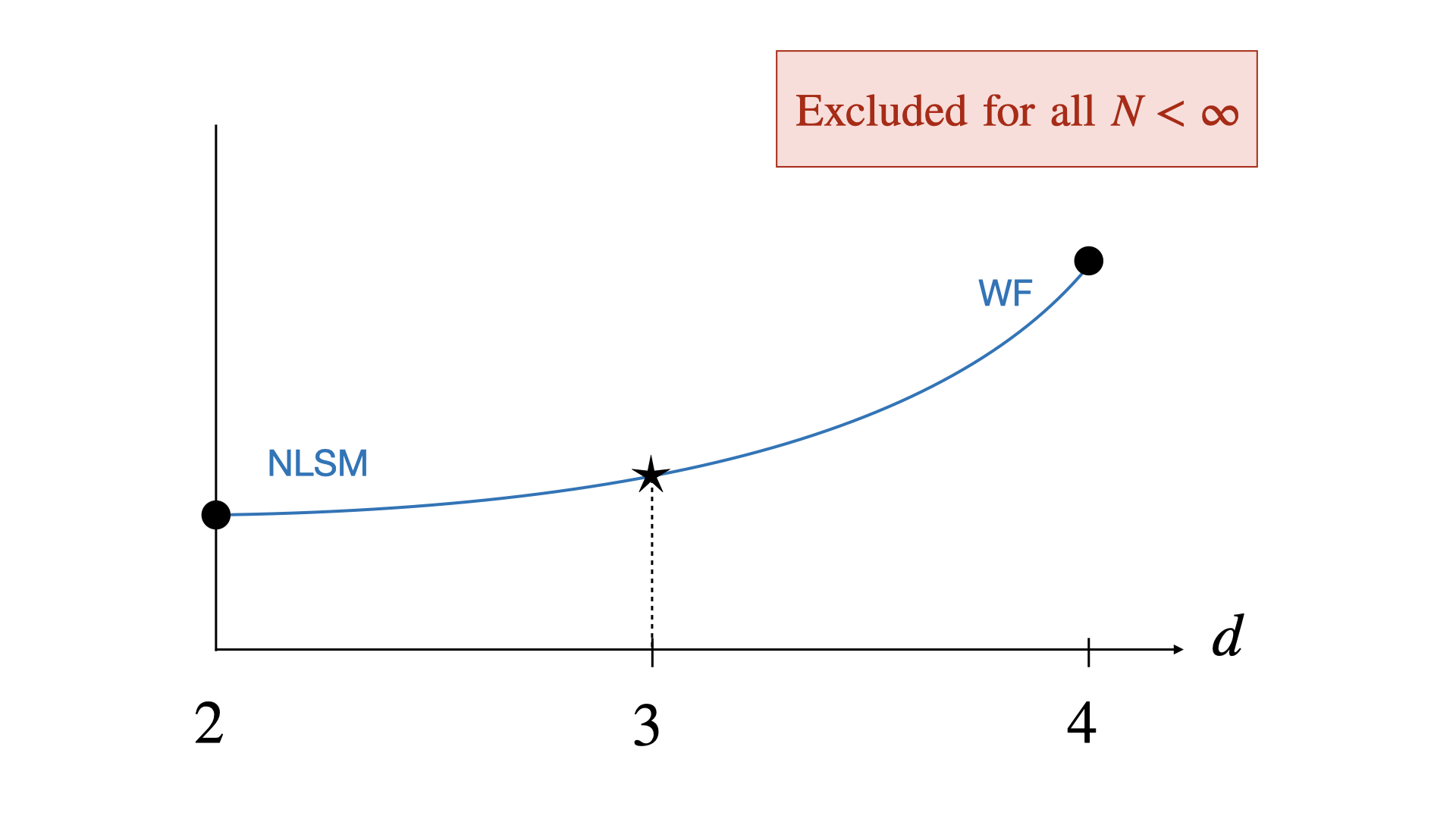} 
	\caption{\textbf{Analytic Connection.} The WF $O(N)$ CFT and the NLSM $O(N)$ CFT are analytically connected in the region $2<d<4$. This is excluded for all finite $N$, but is likely true in the Large-$N$ limit. $\star$ denotes 3D $O(N)$ universality class.}
	\label{fig:scenarioA}
\end{figure}

We consider the $O(N)$ Non-Linear Sigma Model, with the action \eqref{eq:action} which we copy here:
\begin{equation}
S=\frac{1}{2t} \int d^{d}x (\partial_\mu  n^a)^2,
\label{eq:action}
\end{equation}
where the $N$ fields $n^a$, $a=1\dots N$, are constrained to live on the unit sphere, 
\be
n^an^a=1\,. \label{eq:constraint}
\ee
As is well known, in $d=2$ this theory is asymptotically free in the UV \cite{Polyakov:1975rr}, while in the IR the dimensionless coupling $t$ grows and drives the theory to a massive phase. In $d=2+\epsilon$ the coupling $t$ is dimensionful and its classical dimension makes it grow towards the UV. Balancing this classical growth against the quantum one-loop term, one obtains a nontrivial weakly coupled UV fixed point in $d=2+\epsilon$ dimensions. This is what we call here the family of NLSM $O(N)$ CFTs in $d>2$. These CFTs have $O(N)$ invariance with one relevant singlet perturbation (at least for small $\epsilon$). For a long time, it was believed to coincide with the WF $O(N)$ CFT family, obtained by analytically continuing from $4-\epsilon$ dimensions. In other words, a smooth trajectory of CFTs interpolating between the two regimes was expected to exist in the interval $2 < d < 4$, see Fig.~\ref{fig:scenarioA}. This is the Analytic Connection scenario.

However, this interpretation faces a serious obstruction.
We consider the following operator, $O(N)$ pseudoscalar with $N-1$ antisymmetric Lorentz indices \cite{Jones:2024ept}:
\be
  B_{\mu_1\dots\mu_{N-1}}=\epsilon_{a_1\dots a_N} \partial_{[\mu_1} n^{a_1} \partial_{\mu_2} n^{a_2}\dots \partial_{\mu_{N-1}]} n^{a_{N-1}}n^{a_N}\,.
  \label{eq:B_N>3}
 \ee
This operator satisfies the antisymmetrized conservation equation:
 \be
\partial_{[\mu_1}B_{\mu_2\dots\mu_N]}=0\,.
 \label{eq:conservation_N>3}
 \ee
In other words, the differential form $B_{\mu_1\dots\mu_{N-1}}dx^{\mu_1}\wedge \dots dx^{\mu_{N-1}}$ is closed, being the pullback of the volume form from the sigma-model target manifold \cite{Jones:2024ept}. We can also give a pedestrian proof of \eqref{eq:conservation_N>3}. Namely, we need to show that
\begin{equation}
	\epsilon_{a_1\dots a_N} \partial_{[\mu_1} n^{a_1} \partial_{\mu_2} n^{a_2}\dots \partial_{\mu_{N}]}n^{a_N}=0\,.
	\label{eq:O31}
\end{equation}
We differentiate the constraint in Eq.~\eqref{eq:constraint}, and obtain  
\be
n^a\partial_\mu n^a=0\,.
\ee
This equation says that the $N$ vectors $\partial_\mu n^1\dots \partial_\mu n^{N}$ are linearly dependent, implying
\be
\partial_{[\mu_1} n^{1}\partial_{\mu_2} n^{2} \dots \partial_{\mu_{N}]}n^{N}=0\,,
\ee
and in particular \eqref{eq:O31}. Importantly, this argument works in any $d$, integer or not, since we are not using the explicit values of indices $\mu$. We are just using that $\partial_{\mu}v \partial_{\nu}v$ should vanish if antisymmetrized in $\mu$,$\nu$, for any function $v(x)$, and in particular for $v=n^a$.

Another important fact is that the operator $B_{\mu_1\dots\mu_{N-1}}$ is a primary. Indeed it is easy to check that it cannot be written as a derivative of any other local operator built out of $n^a$.

 Some comments are in order. The operator $B$ vanishes in integer $d$ if $N-1>d$ because of antisymmetrization. For example it vanishes for $d=2$, $N\ge 4$ and for $d=3$, $N\ge 5$. But for non-integer $d$ we do not have such constraints because we cannot assign numerical values to indices and count them. In fact it is known from the time of Wilson that non-integer $d$ is in this sense like infinite $d$ \cite{Wilson:1972cf}. So, for non-integer $d$ operator $B$ is nonzero for any $N$. Operators which exist for any noninteger $d$ but vanish for some integer $d$ are called evanescent. If one is interested only in integer $d$, one can usually forget about evanescents.\footnote{Evanescent operators first arose in four-dimensional gauge theories, when considering the renormalization of four-fermion operators in dimensional regularization \cite{Dugan:1990df,Buras:1990fn,Herrlich:1994kh}. In dimensional regularization, evanescents mix under renormalization with physical operators and thus cannot be neglected in perturbative computations. So in those contexts evanescents are important even for describing physics in integer physical dimensions.} However when interpolating between different integer $d$'s, as we are trying to do here, we should definitely pay attention to them. If one $d$-dependent family of CFTs (e.g.~NLSM) has an evanescent operator with some dimension, while another (e.g.~WF) does not have it, these cannot be the same families. And if these are not the same families then other observables, like dimensions of non-evanescent operators, are also expected to be different. 
 In the context of dimensional continuation of scalar theories, evanescent operators were previously discussed in e.g.~\cite{Hogervorst:2014rta,Hogervorst:2015akt}.\footnote{More recent references include  \cite{DiPietro:2017vsp,Ji:2018yaf,Ji:2018emi,Jin:2023cce}.}

\begin{remark}\label{rem:sym}
	Since operator $B_{\mu_1\dots\mu_{N-1}}$ is conserved, one may ask if it corresponds to any symmetry. For example, when we Hodge-dualize $B$ (which is possible in integer $d$) we get a conserved current in $d=3$ for $N=3$, and a conserved current in $d=4$ for $N=4$. So those are 0-form symmetries, using the modern terminology \cite{Gaiotto:2014kfa}. For general $d$, these operators should be understood as $a$-form symmetries where $a$ is non-integer or even negative depending on $d$ and $N$. Objects charged under these symmetries would have a non-integer or negative number of dimensions $a$. Charged objects of fixed integer codimension whose dimension is non-integer, interpolating between say dimension 0 (local operators) and 1 (line operators), are sometimes considered in the context of $4-\eps$ expansion, see e.g. \cite{Chester:2015wao}. Charged objects of negative dimension would be more exotic. For $d=2$ and $N=3$, the two-form $B_{\mu\nu}$, once contracted with $\eps^{\mu\nu}$, corresponds to the $\theta$-term density in the NLSM. The terminology "(-1)-form symmetry" is indeed sometimes used in modern literature with respect to such terms, whose integral is the instanton number \cite{Cordova:2019jnf}. We want to stress that in this work we do not add the $\theta$-term or any other term constructed from operators $B_{\mu_1\dots\mu_{N-1}}$ to the NLSM action. In other words, we view these local operators as a part of the theory, not as a means to perturb the theory.\footnote{We thank Riccardo Argurio and Giovanni Galati for a discussion.}
\end{remark}

As is well known, CFT primary operators satisfying conservation conditions have protected scaling dimensions. In particular, being a closed $N-1$-form, the operator $B$ has a protected dimension equal to $N-1$. To prove this, one could 
start with the two-point function of a generic primary with dimension $\Delta$ and $N-1$ antisymmetric Lorentz indices, differentiate it, and impose the constraint \eqref{eq:O31}. This is easy for $N=3$, but becomes tedious as 
$N$ increases and it is instead more convenient to follow an argument based on the application of commutation relations of the generators of the conformal algebra and the operator, see App.~\ref{app:closedform_dim}.  
So we have a prediction that the scaling dimension of $B$ has to be exactly equal to $N-1$ in $2+\epsilon$ dimensions. 
This is not completely trivial because the classical scaling dimension of $n^a$ being $\epsilon/2$, the classical scaling dimension of $B$ gets corrected at order $\epsilon$, and the anomalous dimension should come out exactly opposite to compensate. As a check, we can compute the one-loop anomalous dimension of the operator $B$ for $N=3$ at the UV fixed point and verify that the dimension is indeed protected, see Appendix \ref{sec:andimB}. As expected, the anomalous dimension exactly cancels with the classical contribution of order $\epsilon$, confirming that the operator has indeed dimension 2 in this case. 

The existence of the protected operator $B$ is an obstruction for the Analytic Connection scenario. The above argument shows its existence in $d=2+\epsilon$, but if we look from the side of the $4-\epsilon$ expansion the situation changes. Indeed, in the $4-\epsilon$ expansion the lowest $O(N)$ pseudoscalar with $N-1$ antisymmetric Lorentz indices is 
\be\label{eq:lowestB4eps}
\epsilon_{a_1\dots a_N}\partial_{[\mu_1}\phi^{a_1}\dots\partial_{\mu_{N-1}]}\phi^{a_{N-1}} \phi^{a_N}\,,
\ee 
of dimension $2N-1+O(\epsilon)$. While this operator has a similar structure to $B$, it is not conserved because the scalar fields are unconstrained in this case.

 As discussed before Remark \ref{rem:sym}, we consider operator $B$ an inseparable part of the theory in $d = 2 + \epsilon$ and in all non-integer $d$ (even though it vanishes in integer $d<N-1$). We then expect it to have nonzero correlation functions with other fields. One simple correlation function to consider is
\be
\langle n^{a_1}(x_1) n^{a_2}(x_2)\dots n^{a_N}(x_N) B_{\mu_1\dots \mu_N}(x_{N+1}) \rangle\,.
\label{eq:corrB_N}
\ee
At tree level (see App.~\ref{sec:andimB} for $N=3$), this correlator is proportional to
\beq
\epsilon^{a_1 a_2\ldots a_N} {p_1}_{[\mu_1} \cdots {p_N}_{\mu_N]}\,.
\eeq
This tensor structure vanishes for integer $d<N-1$, because there are not enough momentum indices to antisymmetrize. But for any non-integer $d$ it is a non-trivial tensor structure. When interactions are turned on, we expect that the prefactor will be corrected to $1+O(\eps)$, but this will not make the correlator vanish. So, we find it hard to believe that operator $B$ may somehow decouple in non-integer $d$. 
Then, the existence of the conserved operator $B$ in $d=2+\epsilon$ and its absence in $d=4-\epsilon$ rules out the scenario in Fig.~\ref{fig:scenarioA} for all finite values of $N$. 

Let us discuss next the large-$N$ limit. In this limit the operator $B$ becomes infinitely heavy and decouples, allowing the two theories to match without obstruction. Since we want to be maximally optimistic, we find it reasonable to believe that the Analytic Connection scenario may be true at large $N$ to all orders in the $1/N$ expansion. 
This expectation is actually consistent with well-established results in the literature. Indeed, in the large-$N$ limit, to all orders in the $1/N$ expansion and within the range $2 < d < 4$, there exists an analytic family of CFTs,  which we denote by "Large-$N$ $O(N)$ CFTs". This family of theories is described by the IR fixed point reached with the Hubbard–Stratonovich action \cite{Moshe:2003xn}
\begin{equation}
S = \int d^d x \left( \frac{1}{2} (\partial_\mu \vec\phi)^2 + \frac{1}{2\sqrt{N}}\vec\phi^2 \sigma - \frac{3}{2 N \lambda} \sigma^2 \right)\,.
\label{eq:HS_action}
\end{equation}
This action can be used to compute perturbative expansions in $1/N$, valid in the whole range $2<d<4$. It is then possible to expand these results around $d = 2 + \epsilon$ and $d = 4 - \epsilon$, and compare with the NLSM and WF $O(N)$ CFTs, when one first computes at small $\epsilon$ and then takes the large-$N$ limit. In both cases, the agreement is exact to all currently known orders in the expansions, providing strong evidence that all three descriptions correspond to the same theory in the large-$N$ limit, presumably to all orders in $1/N$.\footnote{\label{note:largeN}Note that for the NLSM $O(N)$ CFT, the $2+\eps$ expansions have been currently computed completely independently only up to two loops. Higher-order terms are also sometimes available, but they have been obtained using, among other things, agreement with the Large-$N$ $O(N)$ CFT as an input. So those higher-order terms do not provide a fully independent check of the equivalence between the two theories in this limit.}  
 
 There exists also a formal proof of the equivalence of the NLSM CFT in the large-$N$ limit with the Large-$N$ $O(N)$ CFT \cite{Moshe:2003xn}. The proof is based on manipulations of path integrals and does not directly connect to the fact that the NLSM CFT should be defined by analytically continuing from $d=2+\epsilon$. Another proof clarifying this issue would be welcome. 
 
 Although we have already excluded the Analytic Connection scenario without doing any computations, it is worthwhile to review the numerical status of the $2+\epsilon$ in reproducing the 3D critical exponents.
Let us consider $N=3$ first. In this case, numerical results obtained with the NLSM $O(N)$ CFT in $d=2+\epsilon$ show quite poor agreement with known values of the CFT data of the 3D $O(N)$ universality class, especially compared to the very good estimate obtained with the $4-\epsilon$ expansion of the WF $O(N)$ CFT.  In Tab.~\ref{tab:operator_comparison}, we compare the 3D scaling dimensions computed using the conformal bootstrap with the results obtained from the $4-\epsilon$ expansion and the $2+\epsilon$ expansion. In general, the $4-\epsilon$ expansion is known to much higher order than the $2+\epsilon$ series. Nevertheless, for a fair comparison, we apply the same approximation procedure: we truncate both series at four loops and then perform a Padé resummation of the truncated series.\footnote{As already mentioned in footnote \ref{note:largeN}, the $2+\eps$ expansion terms beyond two loops involve large-$N$ input, but we trust them for this exercise.} Actually, it does not matter how one slices and dices the series --- for all procedures we tried, the disagreement between NLSM and the other two columns is simply embarrassing.

We can repeat this analysis for larger $N$. The disagreement gradually gets smaller, and for $N\gtrsim 10$ one gets reasonably good agreement. This is probably in the large-$N$ region where as we argued above NLSM may well be correct to all orders in $1/N$.
 
 \begin{table}[h]
 	\centering
 	\renewcommand{\arraystretch}{1.3}
 	\begin{tabularx}{0.85\textwidth}{XX XX XX}
 		\toprule
 		\multicolumn{2}{l}{\bfseries NLSM $2+\epsilon$} & \multicolumn{2}{l}{\bfseries WF $4-\epsilon$} & \multicolumn{2}{l}{\bfseries Bootstrap} \\
 		\cmidrule(lr){1-2} \cmidrule(lr){3-4} \cmidrule(lr){5-6}
 		Operator & $\Delta$ & Operator & $\Delta$ & Operator & $\Delta$ \\
 		\midrule
 		$n^a$ & $ {0.6}$ & $\phi^a$ & ${0.517}$ & $\varphi$ & $0.5189$ \\
 		$(\partial n)^2$ & ${1^*}$ & $\phi^2$ & ${1.566}$ & $S$ & $1.5949$ \\
 		$O_4$ & $2^{**}$ & $(\phi^2)^2$ & ${3.661}$ & $S'$ & $3.7668$ \\
 		$n^{\{a} n^{b\}}$ & ${1.603}$ & $\phi^{\{a} \phi^{b\}}$ & ${1.211}$ & $t$ & $1.2095$ \\
 		$n^{\{a}n^{b}n^{c} n^{d\}}$ & ${5.104}$ & $\phi^{\{a}\phi^{b}\phi^{c} \phi^{d\}}$ & ${2.983}$ & $\tau$ & $<2.990$ \\
 		\bottomrule
 	\end{tabularx}
 	\caption{Comparison of operator dimensions in 3D for the NLSM, the WF and the Bootstrap $O(3)$ CFTs. The numerical values for the former two methods were obtained by Padé approximating the $2+\epsilon$ and the $4-\epsilon$ series respectively truncated at four loops. We used the [2/2] Padé approximant for all quantities except for $*$, for which we used the [0/4] to avoid poles, and $**$, as only two perturbative orders are known in this case. The $\eps$-series were obtained using the \texttt{Mathematica} package attached to~\cite{Henriksson:2022rnm}, which contains a comprehensive set of CFT data for the critical $O(N)$ models. For details and references regarding the computations of the various expansions, we refer the reader to that work.}
 	\label{tab:operator_comparison}
 \end{table}

We briefly discuss another test. It is common to assume that WF $O(N)$ CFT can be analytically continued down to $d=2$. We can then impose the boundary condition at $d=2$ when doing Pad\'e of scaling dimensions in $4-\eps$ expansions. We can then analyze the \emph{slope} of the resummed expansion around $d=2$, and compare it with what $d=2+\eps$ expansion predicts. For $N=3$, we get very different slopes, suggesting that the curves differ to first order in $d-2$. For larger $N$ the disagreement is again smaller.

\begin{figure}[h]  
	\centering
	\includegraphics[width=0.6\textwidth]{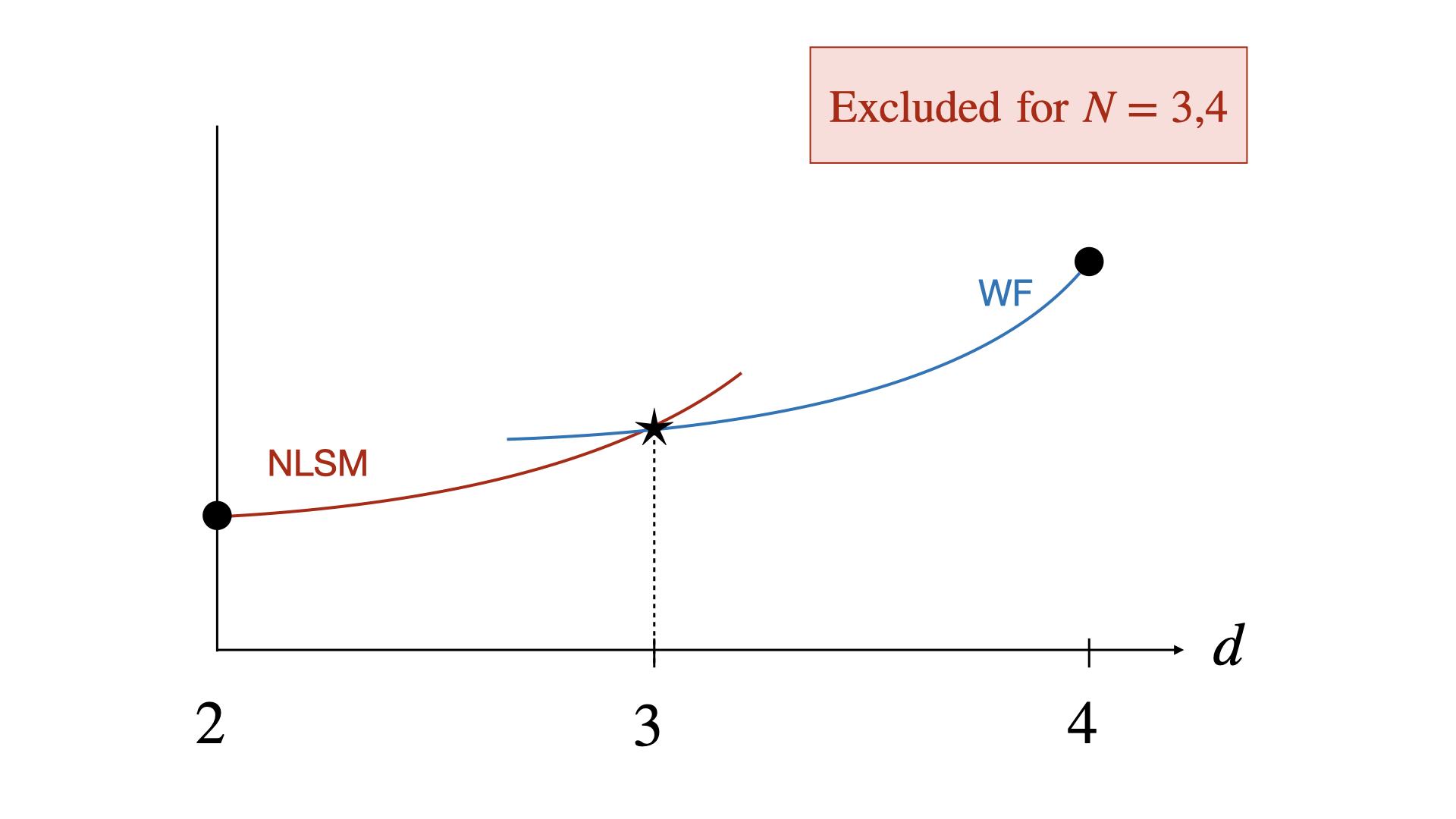}  
	\caption{\textbf{3D Intersection.} The WF $O(N)$ CFT and the NLSM $O(N)$ CFT follow different paths in theory space in the interval $2<d<4$, but cross at $d=3$.  $\star$ denotes 3D $O(N)$ universality class.}
	\label{fig:scenarioB}
\end{figure}

Anyways, since the Analytic Connection scenario has been excluded for all finite $N$, we are interested in investigating possible alternatives. 
One possibility ("3D Intersection") is that the WF $O(N)$ CFT and the NLSM $O(N)$ CFT, while being different theories for $d\ne3$, describe the same fixed point at $d=3$, corresponding to the 3D $O(N)$ universality class, see Fig.~\ref{fig:scenarioB}. 
For a reason similar to the previous scenario, this is allowed only if the operator $B$ vanishes at this specific value of $d$. Let us start by considering $N=3$. In this case, the operator $B$ does not vanish and instead corresponds to a conserved axial current after Hodge-dualization. Clearly, this current is absent in the critical 3D Heisenberg model, which has only $O(3)$ global symmetry and no additional $U(1)$ which could correspond to such an additional current. The situation is similar for $N=4$, as the operator $B$ corresponds to a pseudoscalar with dimension $3$ in this case.\footnote{Note that we infer the existence of this protected pseudoscalar in $d=3$ by continuity, i.e. we run the argument of App.~\ref{app:closedform_dim} for non-integer $d$ where it applies and gives that $B$ has dimension 3 for any such $d$, and then take the limit $d\to 3$. We cannot run the argument of App.~\ref{app:closedform_dim} directly in $d=3$ because $B$ is a top-form in this case and the argument does not apply. We thank Yu Nakayama for a remark prompting us to clarify this.} Such a pseudoscalar is not known to exist in the 3D $O(4)$ universality class. 
This scenario is then excluded for $N=3,4$. The situation changes for larger values of $N$. Indeed, for $N>4$ the operator $B$ becomes evanescent in $d=3$. Therefore, there is no contradiction for the two theories to coincide at the physical value $d=3$ in this range of $N$. So this scenario is in principle allowed for $N>4$, although we consider it rather exotic.

\begin{figure}[h] 
	\centering
	\includegraphics[width=0.6\textwidth]{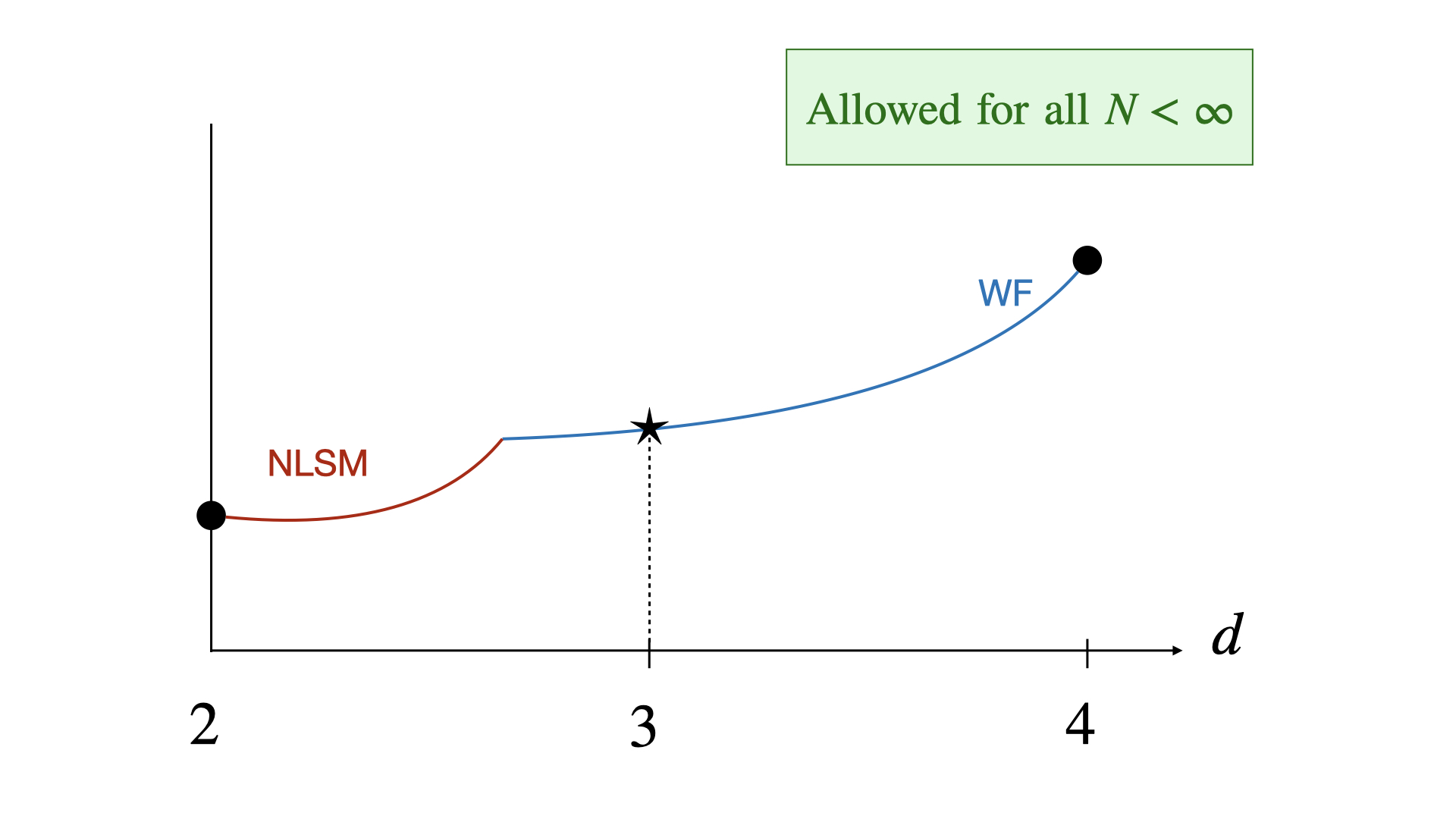}  
	\caption{\textbf{Continuous Connection.} Multiplet recombination occurs (for $2<d<3$ if $N=3,4$), and the WF $O(N)$ CFT and the NLSM $O(N)$ CFT turn out be continuously but not analytically connected. $\star$ denotes 3D $O(N)$ universality class.}
	\label{fig:scenarioC}
\end{figure}

 There exists a third scenario, which would still allow for a connection between the two theories.
 What indeed might happen is that the operator $B$, while protected at small $\epsilon$, becomes unprotected and its dimension lifted for $\epsilon\ge \epsilon_0$. However, such a scenario---that a protected operator becomes lifted---is tightly constrained as the only way it may happen is through multiplet recombination. In this case, the two CFTs would be connected continuously but not analytically, see Fig.~\ref{fig:scenarioC}, where the degree of non-analyticity is exaggerated to make the point clear. Multiplet recombination should arrive for $\eps_0<1$ for $N=3,4$, so that in 3D we do not have any unwanted protected operators. In $N>4$ the protected operator is evanescent in 3D, and one may contemplate recombination for any $\eps_0<2$.
 
Admittedly, multiplet recombination is a rather pathetic attempt to save the $2+\eps$ expansion. Since the critical exponents vary only continuously across the recombination dimension, it would be pretty much impossible to compute the 3D exponents from the $2+\eps$ series. Still, as a theoretical concept, multiplet recombination is an intriguing phenomenon. This possibility is investigated in the next section.

\section{Possibility of recombination}
\label{sec:recomb}
\begin{figure}[h!]  
	\centering
	\includegraphics[width=0.6\textwidth]{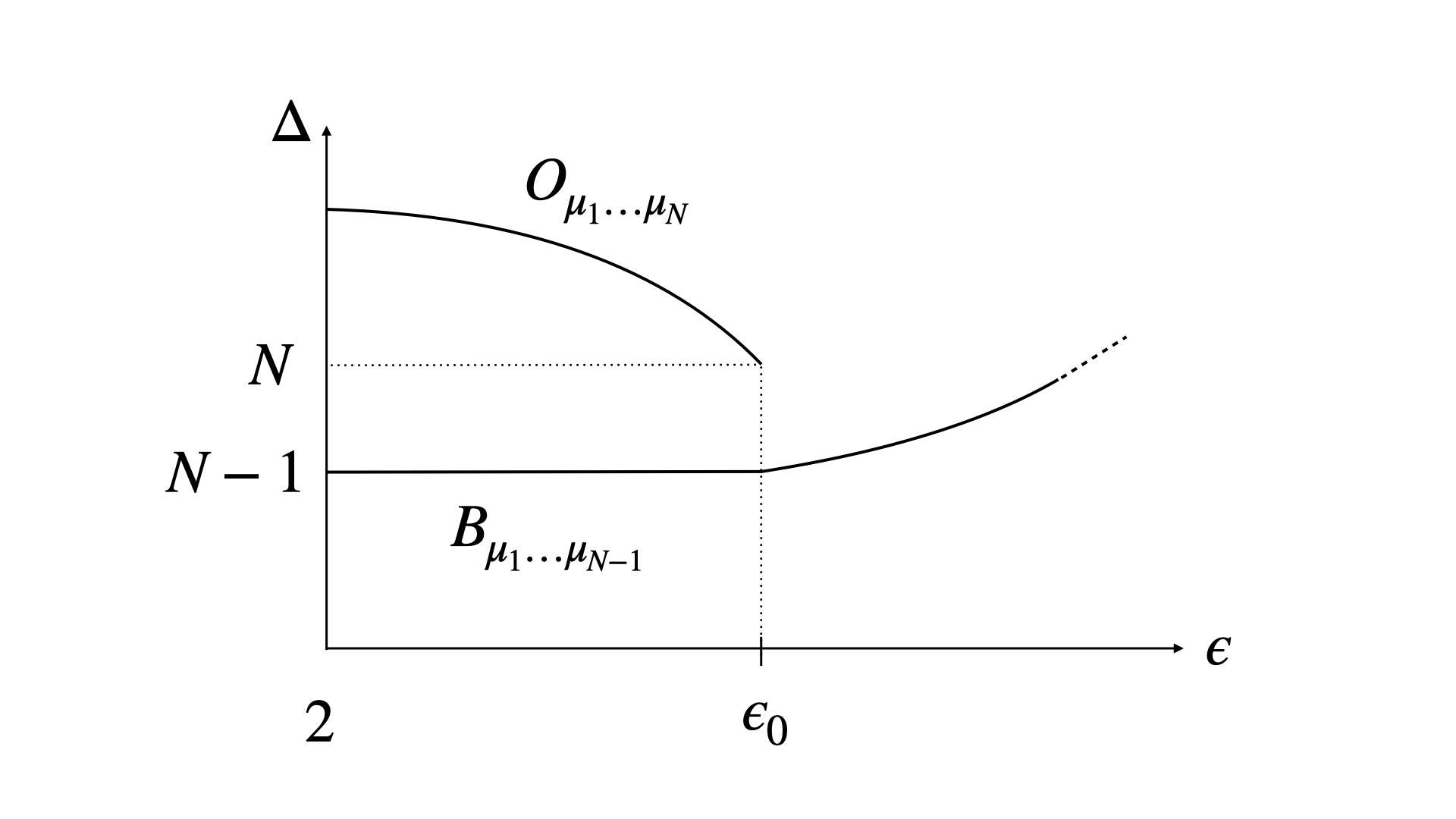}  
	\caption{Multiplet recombination scenario for the operator $B$.}
	\label{fig:recomb}
\end{figure}

The conservation equation \eqref{eq:conservation_N>3} is an example of a shortening condition, saying that the conformal multiplet of $B_{\mu\nu}$ contains fewer states than a multiplet of an antisymmetric primary of a generic dimension $\Delta\ne 0$. If the operator $B$ is to be lifted, the missing states must come from somewhere. It is believed that this can only happen through multiplet recombination (see Section \ref{sec:examples-of-recombination-in-physics} below for various examples). Namely, the missing states must come from the multiplet of another primary $O$.  The quantum numbers of $O$ should be related to those of $B$. Under $O(N)$, it should transform in the same way as $B$, i.e.~be a pseudoscalar. Under rotations, it should transform as $\partial_{[\mu_1} B_{\mu_2\dots \mu_N]}$ i.e.~have $N$ antisymmetric Lorentz indices. Moreover, at the recombination point $\epsilon=\epsilon_0$, the scaling dimension of $ O_{\mu_1\dots\mu_N} $ should equal to the dimension of putative $\partial_{[\mu_1} B_{\mu_2\dots \mu_N]}$ states that it should supply, i.e.~$\Delta_O=\Delta_B+1=N$. For small $\epsilon$, we will find that all candidate $O$'s have a higher dimension than what is necessary for recombination, and thus $B$ has no choice but to stay protected.  Note that this is a non-perturbative statement. But perhaps one $O$ comes down as $\epsilon$ is increased, reaching $\Delta_O = N$ at some $\epsilon=\epsilon_0$. This would allow the operator $ B $ to acquire an anomalous dimension at $\epsilon>\epsilon_0$, while $O$ would disappear from the CFT spectrum, see Fig.~\ref{fig:recomb}. Then, starting from $\eps=\eps_0$, the scaling dimension of $B$ could keep rising and join at $d$ close to 4 the dimension of the operator \eqref{eq:lowestB4eps}. This is how the "Continuous Connection" scenario in Fig.~\ref{fig:scenarioC} can be realized. 

\subsection{Search for candidates for the recombination}

To identify candidates for the recombination, we need to consider the lightest primary operators in the spectrum which are $O(N)$ pseudoscalars with $N$ antisymmetric Lorentz indices. Because of Eq.~\eqref{eq:O31}, the lightest such nonzero operators must have at least $N+2$ derivatives, hence dimension $\Delta=N+2+O(\epsilon)$. It is a bit tedious to go through the list of all possible such operators and figure out which ones are independent.  We refer to App.~\ref{app:list_operators} for the details, and we present here the final result. There exists only one independent such operator with $N+2$ derivatives, which reads
\begin{equation}
(O_{N+2}^A)_{\mu_1\dots\mu_N}=\epsilon_{a_1\dots a_N} \partial_{\ka}\partial_{[\mu_1} n^{a_1} \partial_{\ka}\partial_{\mu_2} n^{a_2}\partial_{\mu_3} n^{a_3}\dots \partial_{\mu_N]} n^{a_N}\,.
\end{equation}
Notably, this
corresponds to a descendant operator, as it satisfies
  \begin{equation}
      (O_{N+2}^A)_{\mu_1\dots\mu_N}=\partial_{[\mu_1} \left(\epsilon_{a_1\dots a_N} \partial_{\ka}n^{a_1} \partial_{\ka}\partial_{\mu_2} n^{a_2}\partial_{\mu_3} n^{a_3}\dots \partial_{\mu_N]} n^{a_N}\right)\,,
  \end{equation}
  and can then be disregarded in our analysis. 

Therefore, the lightest primary operator capable of recombination must have at least $N+4$ derivatives, leading to a dimension $\Delta_{O_{N+4}} = N+4 + \mathcal{O}(\epsilon)$. 
The number of operators with $N+4$ derivatives is significantly larger than that of operators with $N+2$ derivatives, making it more challenging to identify all the relations among operators and determine the independent primary operators, which are the only ones of interest.  We plan to do this, and to compute anomalous dimensions of these operators, in the future work \cite{WIP}.\footnote{ The first arXiv version of this paper considered the anomalous dimension of the primary operator $O_5^E$ for $N=3$ with 5 derivatives. That operator did not have the correct quantum numbers to participate in recombination, being an $O(3)$ scalar, not a pseudoscalar, and so we removed its discussion. We thank an anonymous referee for pointing this out.} Until this is done, we consider Continuous Connection scenario as potentially allowed for all $N\ge 3$.

Note that, having the candidate operators dimension equal to $N+4 + \mathcal{O}(\epsilon)$, the negative anomalous dimension required for recombination would be rather large. However, there are other cases in physics where such large negative anomalous dimensions were observed, leading to physical effects, see e.g. \cite{Kaviraj:2020pwv}, so a priori this cannot be excluded.

\subsection{Examples of recombination in physics}\label{sec:examples-of-recombination-in-physics} In this section we list several examples where recombination of short multiplets is known to occur, or was previously speculated to occur, to make the reader comfortable with the idea. There are many such examples and our list is not meant to be exhaustive.

Ref.~\cite{Rychkov:2015naa} discussed multiplet recombination between the $\phi$ and $\phi^3$ free theory multiplets when the free scalar theory in $d=4$ is deformed into an interacting WF CFT in $d=4-\epsilon$. During this deformation, infinitely many short multiplets corresponding to higher-spin currents of even spins $\ell \ge 4$ also recombine \cite{Skvortsov:2015pea}. This is as expected, because higher-spin symmetry is not supposed to exist in an interacting CFT \cite{Maldacena:2011jn,Alba:2015upa}. On the other hand the stress tensor of spin 2 remains conserved, and in fact, there is no free theory primary it could recombine with.\footnote{See e.g. \cite[Eq.~(A.6)]{Meneses:2018xpu} whose r.h.s.~does not contain the $\chi_{5,\frac 12,\frac 12}$ character needed for the recombination of stress tensor, but it does contain $\chi_{7,\frac 32,\frac 32}$, $\chi_{9,\frac 52,\frac 52}$, $\chi_{11,\frac 72,\frac 72}$,\ldots\ needed for the recombination of spin 4,6,8,\ldots\ higher spin currents.}

Multiplet recombination for conserved currents is somewhat analogous to the Higgs mechanism for gauge fields, by which massless vector bosons become massive while eating scalars which provide the additional states. This analogy becomes precise in the context of holography, where a conserved CFT current is dual to a gauge field in the bulk, and the conservation condition is broken when the standard perturbative Higgs mechanism (a gauge field, and a charged scalar field acquiring a vev) takes place in the bulk \cite{Porrati:2001db}. 

Another interesting holographic example involves pure Yang-Mills theories in AdS$_4$ with Dirichlet boundary conditions (and no charged matter in the bulk).  At small AdS radius, this gives rise to conserved boundary currents dual to gauge fields. This boundary condition corresponds to a deconfined phase in the bulk and should disappear in the large AdS radius limit.  Multiplet recombination of boundary currents at some critical radius was considered as a potential mechanism for this disappearance, among other mechanisms~\cite{Aharony:2012jf,Copetti:2023sya}.  Recently, Ref.~\cite{Ciccone:2024guw} computed the anomalous dimension of the lightest boundary operator potentially involved in the multiplet recombination of the conserved current, as a function of the AdS radius. On the basis of their result, they reached a conclusion similar to our conclusion for $B_{\mu\nu}$ - that the recombination is disfavored.  The authors of \cite{Ciccone:2024guw} found it more plausible that the Dirichlet boundary condition disappears through merger and annihilation. As a matter of fact, this is also going to end up being our favored scenario for NLSM, see Sec.~\ref{sec:assessment}.

A notable class of CFT families consists of conformal manifolds, which are families of theories related by perturbations with exactly marginal scalars. One may ask if, when moving on a conformal manifold, some conserved currents may become lifted, and how this happens. Such examples abound. For example, the compactified scalar boson 2d CFT has generically $U(1)\times U(1)$ symmetry, enhanced to $SU(2)\times SU(2)$ at the self-dual radius. Moving away from the self-dual radius, the extra currents get broken through multiplet recombination.\footnote{At the self-dual point we have 9 marginal scalars $J_i\bar J_j$ where $J_i$, $\bar J_j$ are the $SU(2)\times SU(2)$ currents. Moving away, 8 of these get eaten by the broken currents, and only one marginal scalar remains. See e.g.~\cite{Ginsparg:1987eb}.}

For conformal manifolds of 4d supersymmetric theories, multiplet recombination of conserved currents was discussed in \cite{Green:2010da}. There, it was shown that the conformal manifold may contain special points where there are scalar operators that are marginally irrelevant but not exactly marginal, and when one moves away from such a point on the conformal manifold, these scalars recombine with some conserved currents so that global symmetry is reduced. Similar supersymmetric examples in $d=3$ were discussed in \cite{Baggio:2017mas}.\footnote{We thank Nikolay Bobev for a discussion.}

 \subsection{Comments on non-integer $N$}\label{sec:comments-on-non-integer-n}
 
It is often the case that theories in physics are analytically continued in the number of field components ($N$ in our case) as in the number of spacetime dimensions $d$. Sometimes this procedure is well defined (see e.g.~\cite{Binder:2019zqc}), in other cases we just follow the nose hoping that the future will put everything on solid ground. One of the rules that one needs to follow for non-integer $N$, as for non-integer $d$, is that one cannot use the Levi-Civita tensor for non-integer number of components. On the other hand, one can antisymmetrize indices. The number of indices is always an integer. We followed these rules in \eqref{eq:B_N>3} and \eqref{eq:lowestB4eps}.

If we want to extend our story to non-integer $N$, we have to deal with several issues:
\begin{enumerate}
	\item[Q1.] Are the theories still well-defined?
	\item[Q2.] Which operators can we consider?
	\item[Q3.] Does NLSM still have extra conserved operators?
	\end{enumerate}
	As for Q1, the answer seems to be yes for WF. We are less sure about NLSM, because of the $n^an^a=1$ constraint. But let us be optimistic and assume NLSM also exists.
	
	As for Q2, because of the absence of Levi-Civita, we lose  pseudoscalar operators \eqref{eq:B_N>3} and \eqref{eq:lowestB4eps}. We can still consider their antisymmetrized counterparts. E.g. instead of \eqref{eq:B_N>3} we could consider:
\be
B_{\mu_1\dots\mu_{k-1}}^{a_1\dots a_k}= \partial_{[\mu_1} n^{[a_1} \partial_{\mu_2} n^{a_2}\dots \partial_{\mu_{k-1}]} n^{a_{k-1}}n^{a_k]}\,,
\label{eq:B_N>3nonint}
\ee
where $k$ is an integer. This operator exists for any $N$. For $N=k$ it is conserved by the given arguments. For non-integer $N$ we do not get any extra conserved operator, and the answer to Q3 is no. 

To summarize, we limit ourselves to integer $N$ here, because we are not sure if NLSM makes sense for non-integer $N$, and even if it makes sense, there does not seem to be any extra conserved operator compared to WF, so we have no story to tell.

\section{Last scenario and conclusions}
 \label{sec:assessment}

So far, we considered three scenarios for the relation between the WF $O(N)$ CFT and the NLSM $O(N)$ CFT in the range $2 < d < 4$. The first, widely believed until now, assumes an analytic connection between the two theories, but this is ruled out for finite $N$ due to the presence of a protected operator $B$ in the NLSM CFT which is absent in the WF CFT. (At large $N$, this obstruction disappears, as this operator becomes infinitely heavy.) The second scenario is that the two theories coincide only at $d = 3$, describing distinct fixed points in the rest of the interval. This is excluded for $N = 3,4$ again for the presence of the operator $B$, but is instead allowed (though exotic) for larger $N$, as the operator $B$ is evanescent in this case. The third scenario involves multiplet recombination, leading to a continuous, but non-analytic, connection between the two CFT families.  The results obtained in the last section are not yet sufficient to draw definitive conclusions about this scenario, although we hope to address this in future work \cite{WIP}.

\begin{figure}[h]  
	\centering
	\includegraphics[width=0.6\textwidth]{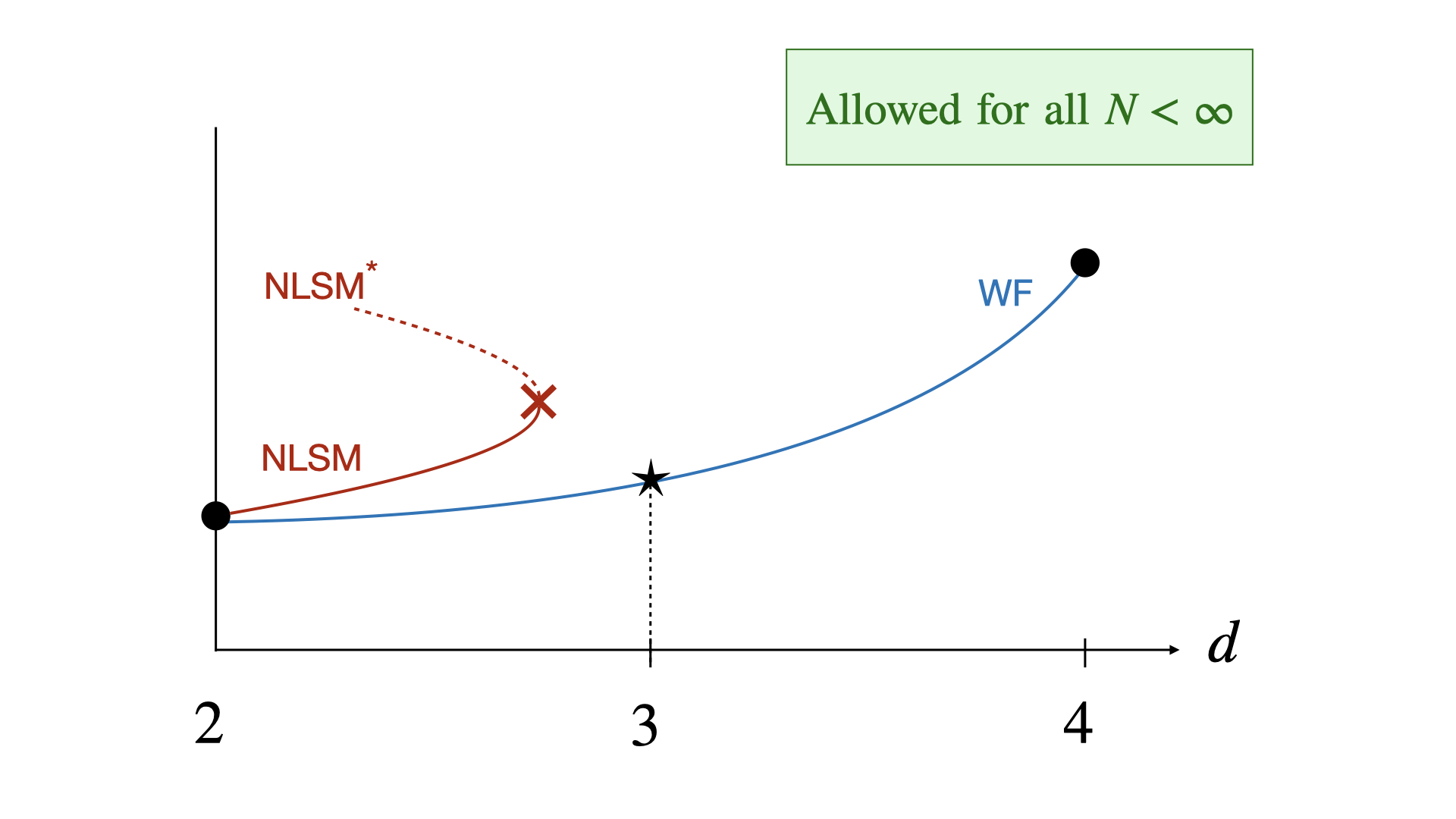}  
	\caption{\textbf{Distinct theories.} The WF $O(N)$ CFT and the NLSM $O(N)$ CFT follow different paths in theory space in the interval $2<d<4$, without crossing. The NLSM $O(N)$ CFT may disappear via merger and annihilation at some dimension $d_m$, either larger or smaller than 3.  $\star$ denotes 3D $O(N)$ universality class.  }
	\label{fig:scenarioD}
\end{figure}

There exists a last scenario, for which the WF $O(N)$ CFT persists analytically down to $d=2$, but follows a path that does not intersect that of the NLSM $O(N)$ CFT, see Fig.~\ref{fig:scenarioD}. This scenario is in principle allowed for all finite values of $N$ and in particular for $N=3$.\footnote{We draw the two families NLSM and WF with a different first derivative at $d=2$ in the theory space, because of the test about extracting slope of scaling dimensions that we described in Section \ref{sec:presentation}.}
In fact, there exist several other hints which also support this proposal, and which we summarize here.

The first important fact, which explains the shape of the upper curve in Fig.~\ref{fig:scenarioD}, is that the NLSM $O(N)$ CFT may disappear in the interval $2<d<4$ through the merger and annihilation mechanism \cite{Kaplan:2009kr, Gorbenko:2018ncu}.  Merger and annihilation happens when an operator which is a "singlet scalar", i.e.~a scalar both under $O(N)$ and rotations, crosses marginality. The leading such operator, $(\partial n)^2$, is relevant for $d=2+\eps$ and its $2+\eps$ expansion shows no signs of approaching marginality. On the other hand, the lightest irrelevant singlet scalar operator,\footnote{As presented in \cite{Wegner:1987gu}, there exists a whole family of operators with $2s$ derivatives and a large negative anomalous dimension, at least at one loop. In this work, we focus on the lightest operator in this family, as we expect that for larger $s$, subleading corrections become less negligible, making it more difficult to trust that marginality is actually reached. For a more detailed analysis of the stability issues associated with this family of operators at large $s$, we refer the reader to Ref.~\cite{Derkachov:1997gc}. }
 \begin{equation}
 O_4=(\partial_{\mu}n^a\partial^{\mu}n^a)^2-\frac{2}{N}\partial_{\mu}n^a\partial^{\nu}n^a\partial_{\nu}n^b\partial^{\mu}n^b\,,
 \label{eq:O4}
 \end{equation}
 has dimension \cite{Brezin:1976ap,Wegner:1987gu}
 \begin{equation}
\Delta_{O_4}=4-\frac{2}{N-2}\epsilon+\mathcal{O}(\epsilon^2)\,,
\label{eq:DeltaO4}
\end{equation}
which does reach marginality at $\epsilon_m=2-4/N$, i.e.~at $d<4$.\footnote{We only use the one-loop anomalous dimension result. Ref.~\cite{Brezin:1996ff} derives a two-loop expression for the scaling dimension $\Delta_{O_4}$ which if valid could help in finding a better estimate for the value of $\epsilon$ at which merger and annihilation occurs. However, that result appears to be incorrect, as it does not agree with the corresponding large-$N$ limit. This discrepancy is not surprising, given that the derivation relies on consistency with the results of Ref.~\cite{Castilla_1997}, which also presents issues. In fact, the latter is incompatible with the result of the same computation in Ref.~\cite{Derkachov:1997gc}, which is instead supported by agreement with other large-$N$ results.} So that is where we expect merging with another putative family of CFTs, denoted by a dashed line in Fig.~\ref{fig:scenarioD}. 

Note that the smaller the value of $N$, the lower the merger dimension, and thus the more reliable the perturbative estimate itself on which we are basing our discussion. For $N=3$, taking Eq.~\eqref{eq:DeltaO4} at face value, we will have merger around $d_m=2+\eps_m\approx 2.67$. 

 What is this "dashed" family, which we call NLSM${}^*$ by analogy with QCD${}^*$ in \cite{Kaplan:2009kr}? A priori we do not know. It can be said on general grounds that: the operator that was approaching marginality from the irrelevant side in NLSM does so from the relevant side in NLSM${}^*$; that close to the merger there is a short RG flow from NLSM${}^*$ to the NLSM triggered by this operator; and that the merger curve has a parabolic shape so that all CFT data show square root singularities near the merger point \cite{Gorbenko:2018ncu}.\footnote{That the shape must be parabolic suggests that Eq.~\eqref{eq:DeltaO4} overestimates $\eps_m$. See \cite{Gukov:2016tnp} for similar discussions.} Furthermore, on the other side of the merger, the NLSM and NLSM${}^*$ become a complex-conjugate pair of complex CFTs \cite{Gorbenko:2018ncu}, in which the scaling dimension of $\Delta_{O_4}$ acquires an imaginary part. So in no way can they correspond near $d=4$ to the fixed point obtained via the $4-\epsilon$ expansion, where the scaling dimensions of the light operators are all real.\footnote{In the higher part of the spectrum, some WF dimensions do have imaginary parts, related to the non-unitarity of the WF theory in non-integer $d$ \cite{Hogervorst:2015akt}.}

It appears quite plausible that merger and annihilation happens for $N=3$, where this is perhaps our favorite scenario, also for reasons mentioned below. It may well also happen for $N>3$.

One may wonder which phase transition is then described by the NLSM $O(N)$ CFT.
 As a possible answer to this question, Ref.~\cite{Nahum_2015} proposed that the NLSM $O(3)$ family, instead of reaching the 3D $O(3)$ universality class, describes a variant transition in which hedgehog configurations are suppressed \cite{PhysRevLett.71.1911,Motrunich:2003fz}. This is believed to describe the deconfined critical point of the Néel-VBS transition ~\cite{Senthil:2003eed}. There are several arguments supporting this idea. First, the theoretical description of this transition is believed to be the NCCP$^1$ model  \cite{Senthil:2003eed,Senthil:2004fuw,Senthil_2005}, which in $d=2+\epsilon$ reduces to the  NLSM $O(3)$ model.\footnote{Ref.~\cite{Lawrie:1983aw} shows that the gauge field in the NCCP$^1$ model can be integrated out at $d=2+\epsilon$, leaving a $CP^1$ NLSM model. Since $CP^1=S^2$, this is exactly the theory that we are considering.} Then, the NLSM predictions for critical exponents show better numerical agreement with the lattice results obtained in \cite{Nahum_2015} than with the values of standard critical $O(3)$.  Most importantly, in the context of the main argument discussed in this work, the hedgehog-suppressed critical $O(3)$ model possesses a ``topological'' $ U(1) $ current in $d=3 $ \cite{Motrunich:2003fz}, which would correspond to the Hodge dual of the protected antisymmetric operator $B$ observed in $2+\epsilon$. 
 
 We note that there are still many open questions about the (2+1)D Néel-VBS transition. 
 Even the order of this transition remains unclear. While initially believed second-order, more advanced numerical studies pointed to a weakly first-order phase transition \cite{Nahum_2015,Takahashi:2024xxd}. But why is the first-order transition weak? One natural possibility is that $d=3$ lies above $d_m$ where $O(3)$ NLSM merged and annihilated with NLSM${}^*$. Then in $d=3$ there is no unitary CFT with requisite symmetry, and transition has to be first order.  If $d$ is just above $d_m$, it will be \emph{weakly} first-order, due to slowing down of the RG flow, referred to as "walking", or also pseudo-criticality. This would agree very well with the above discussion. 
 
 On the other hand, other researchers believe that the Néel-VBS transition is weakly first-order due to the vicinity of a tricritical point rather than to walking \cite{Chester:2023njo,Takahashi:2024xxd,chen2024phases,Chen:2024jxe}. So the situation remains murky.
 
 Another aspect of the (2+1)D Néel-VBS transition is that it shows an emergent $SO(5)$ global symmetry, which is an extension of $O(3)\times U(1)$ visible in the $O(3)$ NLSM in $d=3$. An alternative description of the VBS–Néel fixed point that makes this symmetry manifest is by the $SO(5)$ WZW model \cite{PhysRevLett.95.036402,PhysRevB.74.064405}. Refs.~\cite{Ma:2019ysf,Nahum:2019fjw} proposed to reach this latter theory via $d=2+\epsilon$ by analytic continuation of $SU(2)_1$ WZW CFT in $d=2$, finding again evidence of merger annihilation for a $d_m<3$, thus supporting the walking rather than tricritical scenario. This new dimensional continuation is at present not very systematic, compared to the traditional $2+\eps$ expansion for $O(N)$ models that we discussed here, as WZW terms are not well defined in non-integer dimensions.\footnote{See also \cite{He:2020azu} which studied the dimensional interpolation between $SO(5)$ and $SU(2)_1$ WZW models via the conformal bootstrap. See also \cite{Zhou:2023qfi} which studied $SO(5)$ WZW model using the fuzzy sphere regulator and found evidence for the walking scenario.}

 To conclude, in this work we lifted the lid on a problem that had simmered for too long. It is high time to bring it to a boil and finish the cooking! Below, we outline several potential future computations and problems that could help clarify the issues discussed:

\begin{itemize}
\item Computing the $\mathcal{O}(\epsilon^2)$ corrections to $\Delta_{O_4}$ in Eq.~\eqref{eq:DeltaO4} seems like a relatively easy task, which could help us in better investigating the Distinct Theories scenario. This would provide a more accurate estimate of the merger dimension as a function of $N$. In particular, for $N=3$, it could help determine whether the Néel–VBS transition is first or second order. 

\item To further investigate the Continuous Connection scenario, one should identify NLSM primary operators with $N+4$ derivatives and $N$ antisymmetric Lorentz indices which are $O(N)$ pseudoscalars, and compute their anomalous dimensions.

\item As a general request, it would be great to do all new NLSM calculations without imposing the large-$N$ consistency (as is the standard for WF calculations), so that one can later get additional independent checks for this consistency.

\item Throughout this work, we have extracted information from the spectrum of the NLSM $O(N)$ CFT in $2+\epsilon$. Further insights might be obtained from the WF $O(N)$ CFT in $4 - \epsilon$. Specifically, for multiplet recombination - and thus for the Continuous Connection scenario - to occur, the lightest operator in the representation of $B$ must reach the scaling dimension of $B$ at precisely the same value of $d$ at which recombination happens in the NLSM. One could therefore compute the anomalous dimension of the operator in Eq.~\eqref{eq:lowestB4eps}, and examine whether - and at what value of $\epsilon$ - it approaches $N - 1$.

\item
In this work, we dealt with the simplest NLSM. One may be interested in NLSM with other target manifolds $\mathcal{M}$ and in their analytic continuation to $d=2+\eps$ dimensions. In all cases closed forms on $\mathcal{M}$ will after pullback give rise to protected operators.\footnote{Since $H^p(S^{N-1})$ is nontrivial only for $p=N-1$ (and 0), in our case we only had the pullback of the volume form. If other cohomology groups are nontrivial there will be other protected operators. Such operators exist e.g.~for the supersymmetric sigma-models on the Calabi-Yau manifolds, where they correspond to the K\"ahler and complex structure moduli, and to the holomorphic top form \cite{Greene:1996cy}. We thank Nikolay Bobev, Andres Collinucci and Mario De Marco for discussions.}
It would be interesting to identify other situations, similar to the ones discussed in this work, where these protected operators preclude the analytic connection of the NLSM description in $d=2+\eps$ with the "naive" WF description in $d=4-\eps$.

\end{itemize}

    \acknowledgments
    
    SR thanks Max Metlitski and Robert Jones for the communication of their related results reported in \cite{Jones:2024ept}. FDC thanks Rajeev Erramilli and Johan Henriksson for useful discussions. SR is supported in part by the Simons Foundation grant 733758 (Simons Bootstrap Collaboration). SR thanks the International Solvay Institute for Physics (Brussels) for hospitality while this work was being finalized. SR thanks Riccardo Argurio, Nikolay Bobev, Andres Collinucci, Mario De Marco and Giovanni Galati for useful discussions, and Victor Horta for the H\^otel Solvay. We thank Gabriel Cuomo, Yu Nakayama and Cenke Xu for comments on the first arXiv version of the paper which led us to clarify some points. We thank an anonymous referee for an insightful comment.
    
    \appendix
\section{Constructing operators in the NLSM}
\label{app:list_operators}
All operators of the NLSM $O(N)$ CFT are built of $n^a$ and its derivatives. The low-dimension operators are easy to enumerate. However, in this work we need to study operators of relatively high dimension. We need to develop some rules that allow us to list all operators systematically at a given dimension, hopefully without overcounting. So let us present some general rules for enumerating operators. We will be focusing on operators that are scalars or pseudoscalars under the $O(N)$ global symmetry. 

\subsection{General rules for $O(N)$ scalars and pseudoscalars}
We introduce two types of building blocks, corresponding to $O(N)$ indices  contracted either with the Kronecker or with the Levi-Civita tensor, and Lorentz indices left uncontracted:
\begin{equation}
\begin{aligned}
   D_{\bar\mu^{(1)}\bar\mu^{(2)}}&=\partial_{\bar\mu^{(1)}}n^a\partial_{\bar\mu^{(2)}}n^a\,,\\
E_{\bar\mu^{(1)}\ldots\bar\mu^{(N)}}&=
	\epsilon_{a_1\dots a_N}\partial_{\bar\mu^{(1)}}n^{a_1}\ldots\partial_{\bar\mu^{(N)}}n^{a_N}\,.
    \end{aligned}
    \label{eq:blocks}
\end{equation}
Here the barred greek letters stand for groups of Lorentz indices (multi-indices): 
\be
\bar\mu^{(j)} = \{{\mu_1,\ldots,\mu_{k_j}}\},\quad \partial_{\bar\mu}=\partial_{\mu_1}\ldots \partial_{\mu_{k_j}}
\ee
We will construct our operators by multiplying these blocks. Since the product of two Levi-Civitas can be expressed in terms of Kroneckers,  we do not have to consider more than one $E$-block per operator. We will also be contracting (some of) the Lorentz indices within blocks and among different blocks. Note that since we work in a generic non-integer dimension, we are only allowed to contract with $\delta_{\mu\nu}$, as the $\epsilon$-tensors do not make sense in non-integer dimensions.

The resulting operator will be a scalar or pseudoscalar under $O(N)$ depending on whether the number of $E$-blocks $k_E$ is 0 or 1.

By imposing the constraint that fields live on the sphere, we can reduce the number of independent building blocks. Differentiating the constraint $n^a n^a=1$ multiple times, we can express the building block  $D_{\emptyset \bar\mu}$ as a linear combination of $D_{\bar\mu^{(1)}\bar\mu^{(2)}}$ where both $\bar\mu^{(1)}$ and $\bar\mu^{(2)}$ are nonempty. 

We also have the equation of motion (EOM) of the NLSM, which can be written as
\be
\Box n^a=-(\partial_\lambda n^b\partial_\lambda n^b) n^a\,.
\label{eq:EOM}
\ee
Operators differing by something proportional to (a derivative of) the EOM will have equal correlation functions at non-coincident points, so we do not have to distinguish them. Differentiating the EOM, we conclude that any building block containing
\be
 \partial_{\mu_{1}}\dots\partial_{\mu_{m}}\Box n^a
 \ee
can be eliminated in favor of a linear combination of products of blocks not containing $\Box$. This means that, when we construct our operators, we can eliminate all operators where indices are contracted within any of the multi-indices $\bar\mu^{(j)}$.

To summarize, we identified two general rules to eliminate certain building blocks:
\begin{itemize}
	\item
	Blocks $D_{\emptyset\bar\mu}$.
	\item
	Any block containing contracted indices within a string of derivatives acting on $n^a$.
\end{itemize}

Let us now focus on operators with a total number of derivatives equal to $p$. As for every block $E$ we need at least $N-1$ derivatives and for every block $D$ we need at least 2 derivatives, the corresponding numbers of structures $k_E, k_D$ composing the operator must satisfy
\be
(N-1) k_E+ 2 k_D\le p\,,
\label{eq:k_constraint}
\ee
recalling that $k_E=0$ or 1.

\subsection{Operators with antisymmetric open indices}
The antisymmetrization of Lorentz indices further reduces the number of independent operators.  To see this explicitly, let us consider the operators $O$ that have the $O(N)$ and rotation numbers to potentially participate in the recombination scenario described in the main text, i.e.~pseudoscalars ($k_E=1$) with $N$ antisymmetric Lorentz indices. 

We only consider here the case when $O$ has $N+2$ derivatives, which is the lowest number for such operators to exist.  

We start with the case $N = 3$. 
Setting $p = 5$ in Eq.~\eqref{eq:k_constraint}, we find $k_D = 0,1$ for $k_E = 1$. We then list all the corresponding operators, removing building blocks that are dependent due to the rules presented in the previous section, as well as operators that vanish trivially due to the antisymmetrization of open indices. Specifically, we discard operators in which antisymmetrized derivatives act on the same field as well as those containing  $ \partial_{[\mu} v \partial_{\nu]} v $, for any function $v(x)$. We also remove terms proportional to $ \partial_{[\mu}n^{a}\partial_{\nu}n^{b}\partial_{\lambda]}n^{c} $, as they reduce to Eq.~\eqref{eq:O31}. The resulting list is presented in Tab.~\ref{tab:op5N=3}. 

\begin{table}[h]
	\centering
	\ra{1.45}  
	\begin{tabular}{@{}l l l l@{}} 
		\toprule
		$k_E$ & $k_D$ & \textbf{Name} & \textbf{Operator} \\ 
		\midrule
		$1$ & $0$ & $(O_5^A)_{\lambda\mu\nu}$ & $\epsilon_{abc} \partial_{\ka}\partial_{[\mu} n^a \partial_{\ka}\partial_{\nu} n^b \partial_{\lambda]} n^c$ \\  
		1 & 1& $(O_5^B)_{\lambda\mu\nu}$ & $\epsilon_{abc} \partial_{[\mu} n^a\partial_{\ka}\partial_{\nu} n^b  n^c \partial_{\lambda]}  n^d \partial_{\ka} n^d$\\
		1  & 1 & $(O_5^C)_{\lambda\mu\nu}$ & $\epsilon_{abc} \partial_{[\mu} n^a \partial_{\nu} n^b n^c \partial_{\lambda]} \partial_{\ka} n^d \partial_{\ka} n^d$ \\  
		1  & 1 & $(O_5^D)_{\lambda\mu\nu}$ & $\epsilon_{abc} \partial_{[\mu} n^a \partial_{\ka} n^b n^c \partial_{\nu}n^d \partial_{\ka} \partial_{\lambda]}  n^d$  \\
		\bottomrule
	\end{tabular}
	\caption{ Table of $O(3)$ pseudoscalars with 3 antisymmetric Lorentz indices and 5 derivatives in total. $O_5^{B,C,D}$ are proportional to $O_5^{A}$, see the text. }
	\label{tab:op5N=3}
\end{table}

It turns out that this list can be further reduced, as there are relations among the operators that are not captured by the rules we have presented above. There are several ways to see that. Perhaps the simplest way in practice is to replace the expression of the constrained $n^a$ fields with the unconstrained Goldstone bosons $\pi^i$, as explained in App.~\ref{app:review}. Then one can see that operators $O_5^{B,C,D}$ are proportional to $O_5^{A}$.

Below we would like to give a demonstration of this fact working with the $n^a$ fields. While not essential for our purposes, this is instructive, as it illustrates a few mechanisms due to which the rules of the previous section are not complete. Perhaps in the future it may show the way to a more complete set of rules which would allow to write directly a set of linearly independent operators.

The first two identities arise from the fact that 
\be
\partial_{[\mu}n^{a}\partial_{\nu}n^{b}\partial_{\lambda]}n^{c} = 0\,,
\label{eq:vanish}
\ee 
whether or not $a,b,c$ are antisymmetrized or not. We can generate more vanishing quantities multiplying \eqref{eq:vanish} by any function of the fields and applying any derivative. This way we can generate identities among operators which are not immediately obvious. For our purposes, we consider the two identities
\begin{align}
	\partial_\ka \big( \epsilon_{abc} \partial_{[\mu} n^a \partial_{\nu} n^b n^c \partial_{\lambda]}  n^d \partial_{\ka} n^d \big) & = 0\,,\\ 
	 \Box (\epsilon_{abc}\partial_{[\mu}n^a\partial_{\nu}n^b \partial_{\lambda]}n^c)&=0\,.
\end{align}
The first of these implies: 
\begin{equation}
	O_5^C = -2O_5^B \,.
\end{equation}
In the second case, we act with the $\Box$ and use the EOM \eqref{eq:EOM} in the terms where both derivatives fall on the same field. The use of the EOM increases the number of fields $n^a$ by 2. This way we obtain
\begin{equation}
    O_5^A = O_5^C\,.
\end{equation}
The remaining identity, between $ O_B $ and $ O_D $, seems to require a more ad hoc series of manipulations. It can be obtained as follows (see below):
\begin{equation}
\begin{aligned}
    O_B &= \frac{1}{2} \partial_{[\mu} n^1 
    \big( \partial_{\ka}\partial_{[\nu} n^2  n^3 - \partial_{\ka}\partial_{[\nu} n^3  n^2 \big) 
    \big( \partial_{\lambda]}  n^2 \partial_{\ka} n^2 + \partial_{\lambda]}  n^3 \partial_{\ka} n^3 \big) + \dots
    \\
    &=-  \frac{1}{2} \partial_{[\mu} n^1 
    \big( \partial_{\ka} n^2  n^3 - \partial_{\ka} n^3  n^2 \big) 
    \big( \partial_{\nu} n^2 \partial_{\ka} \partial_{\lambda]}  n^2  + \partial_{\nu} n^3  \partial_{\ka} \partial_{\lambda]}  n^3 \big) + \dots =-  O_D \,,
\end{aligned}
\end{equation}
Here the $\dots$ represent antisymmetrized permutations of the indices $1,2,3$. In the first line we could drop $\partial_{\lambda]}  n^1 \partial_{\ka} n^1$ in the second bracket because 
\be
\partial_{[\mu} n^1 \partial_{\lambda]} n^1 =0\,.
\label{eq:adhoc1}
\ee
Similarly in the second line we could drop $\partial_{\nu} n^1 \partial_{\ka} \partial_{\lambda]}  n^1$ since $\mu$ and $\nu$ are antisymmetrized.
Finally, to show that the first and the second lines are equal one has to use the identity $ n^3\partial_\lambda n^3 + n^2\partial_\lambda n^2= -n^1\partial_\lambda n^1$, and then again \eqref{eq:adhoc1}.

 In summary, this shows that the operators $O_5^A,O_5^B,O_5^C,O_5^D$ in Tab.~\ref{tab:op5N=3} are all proportional. We keep $O^A_5$ as the independent operator. 

\begin{table}[h]
	\centering
	\ra{1.45}  
	\begin{tabular}{@{}l l l l@{}} \toprule
		$k_E$ & $k_D$ & \textbf{Name} & \textbf{Operator} \\ 
		\midrule
		$1$ & $0$ & $(O_{N+2}^A)_{\mu_1\dots\mu_N}$ & $\epsilon_{a_1\dots a_N} \partial_{\ka}\partial_{[\mu_1} n^{a_1} \partial_{\ka}\partial_{\mu_2} n^{a_2}\partial_{\mu_3} n^{a_3}\dots \partial_{\mu_N]} n^{a_N}$ \\  
		1& 1& $(O_{N+2}^B)_{\mu_1\dots\mu_N}$ & $\epsilon_{a_1\dots a_N} \partial_{[\mu_1} n^{a_1}\dots \partial_{\mu_{N-2}} n^{a_{N-2}}\partial_\ka \partial_{\mu_{N-1}} n^{a_{N-1}} n^{a_N} \partial_{\mu_N]} n^b \partial_{\ka} n^b$  \\  
		1 & 1 &$(O_{N+2}^C)_{\mu_1\dots\mu_N}$ 
		& $\epsilon_{a_1\dots a_N} \partial_{[\mu_1} n^{a_1}\dots \partial_{\mu_{N-1}} n^{a_{N-1}}  n^{a_N} \partial_{\mu_N]} \partial_{\ka} n^b \partial_{\ka} n^b$ \\  
		1 & 1  & $(O_{N+2}^D)_{\mu_1\dots\mu_N}$ 
		&$\epsilon_{a_1\dots a_N} \partial_{[\mu_1} n^{a_1}\dots \partial_{\mu_{N-2}} n^{a_{N-2}}\partial_\ka n^{a_{N-1}}  n^{a_N} \partial_{\mu_{N-1}} n^b \partial_{\ka}\partial_{\mu_{N}]}  n^b$ \\ 
		\bottomrule
	\end{tabular}
	\caption{Table of $O(N)$ pseudoscalar operators with $N$ antisymmetric Lorentz indices and $N+2$ derivatives for $N\ge 4$. All these operators are proportional to each other, see the text. }
	\label{tab:op5N>3}
\end{table}

Let us now consider the case $N>3$, focusing on operators with $N$ antisymmetric indices and $N+2$ derivatives. All the considerations made for $N=3$ also apply in this case. The constraint in Eq.~\eqref{eq:vanish} generalizes to 
\begin{equation}
    \partial_{[\mu_1}n^{a_1}\dots\partial_{\mu_N]}n^{a_N}=0,
\end{equation} 
as a consequence of Eq.~\eqref{eq:conservation_N>3}. 

 The list of operators is given in Tab.~\ref{tab:op5N>3}; they are all natural generalization of the $N=3$ operators.

By the same procedure as for $N=3$ it is possible to show that the operators $O^A, O^B,O^C,O^D$ are all proportional:  
\begin{equation} 
\begin{aligned} O_{N+2}^A &= \frac{2}{N-1} O_{N+2}^C\,,\\ O_{N+2}^C &= -(N-1) O_{N+2}^B\,,\\ O_{N+2}^D &= -O_{N+2}^B\,. \end{aligned} 
\end{equation}
 We are then left with only one independent operator for all $N\ge3$.

\section{Scaling dimension of a closed differential form}
\label{app:closedform_dim}
The scaling dimension of a primary operator corresponding to a closed differential $p$-form is equal to $p$. To show this, we present a detailed version of the argument outlined in \cite{Jones:2024ept}, based on commutation relations of the generators of the conformal algebra and the operator itself. 

 This result is only valid for $p\ne d$.\footnote{We thank Yu Nakayama for pointing this out, which prompted us to discuss this.} For the top-form $p=d$ the conservation constraint is automatic. Any such top-form is Hodge-dual to a primary scalar, whose scaling dimension is not constrained. We will see in the end why the argument breaks down for $p=d$.

Let us first recall some useful relations. In a CFT, the commutator between the generator of translations $P_\nu$ and the generator of special conformal transformations $K_\mu$ gives a linear combination of the dilatation and the rotation generators, $D$ and $M_{\mu\nu}$ respectively:\footnote{We use commutation relation conventions from \cite{Simmons-Duffin:2016gjk}.}
\begin{equation}
    [K_\mu,P_{\nu}]=2\delta_{\mu\nu}D-2M_{\mu\nu}\,.
\end{equation}
Moreover, given a primary $O$ with conformal dimension $\Delta_O$ inserted at the origin, we have
\begin{equation}
    \begin{aligned}
        &[K_\mu,O(0)]=0\,,\\
        &[P_\nu,O(0)]=\partial_\nu O(0)\,,\\
        &[D,O(0)]=\Delta_O O(0)\,,\\
        &[M_{\mu\nu},O(0)]=\mathcal{S}_{\mu\nu} O(0)\,,
    \end{aligned}
\end{equation}
where $\mathcal{S}_{\mu\nu}$ are the rotation matrices in the $SO(d)$ representation of $O$. 

Let us now consider a primary   $F$ which is a closed $p$-form, i.e. an operator with $p$ antisymmetric indices satisfying
\begin{equation}
    \partial_{[\nu}F_{\mu_1\dots\mu_p]}=0\,,
\end{equation}
or equivalently
\begin{equation}
    [P_{[\nu},F_{\mu_1\dots\mu_p]}(0)]=0\,.
\end{equation}
Being a primary, $F$ also satisfies
\begin{equation}
    [K_{\mu},F_{[\mu_1\dots\mu_p]}(0)]=0\,.
\end{equation}
Using these relations and applying the Jacobi identity to the equation
\begin{equation}
    [K_\mu,[P_{[\nu},F_{\mu_1\dots\mu_p]}(0)]]=0\,,
\end{equation}
we get
\begin{equation}
   [ [K_\mu,P_{[\nu}],F_{\mu_1\dots\mu_p]}(0)]]=0\,
\end{equation}
and therefore
\begin{equation}
    \Delta_F \delta_{\mu[\nu} F_{\mu_1\dots\mu_p]}-\mathcal{S}_{\mu[\nu}F_{\mu_1\dots\mu_p]}=0\,.
    \label{eq:commutationF}
\end{equation}
Now we simply need to replace the explicit expression of the rotation matrices on operators with $p$ antisymmetric indices
\begin{equation}
\mathcal{S}_{\mu\nu}F_{\mu_1\dots\mu_p}=\sum_{i=1}^p\delta_{\nu \mu_i}F_{\mu_1\dots\mu_{i-1}\mu \mu_{i+1}\dots \mu_p}-\sum_{i=1}^p\delta_{\mu \mu_i}F_{\mu_1\dots\mu_{i-1}\nu \mu_{i+1}\dots \mu_p}\,.
\end{equation}
Antisymmetrizing over all indices except for $\mu$, we get
\begin{equation}
\mathcal{S}_{\mu[\nu}F_{\mu_1\dots\mu_p]}=p \ \delta_{\mu[\nu}F_{\mu_1\dots\mu_p]}\,,
\end{equation}
and therefore, replacing in Eq.~\eqref{eq:commutationF}, we obtain
\beq
(\Delta_F-p) \ \delta_{\mu[\nu}F_{\mu_1\dots\mu_p]} =0.
\eeq
This implies $\Delta_F=p$. Unless of course $\delta_{\mu[\nu}F_{\mu_1\dots\mu_p]}$ is identically zero, in which case the equation holds trivially for arbitrary $\Delta_F$. This exceptional case is realized if the number of spacetime dimensions $d$ is an integer and $p=d$. 

\section{The UV fixed point for the NLSM}
\label{app:review}
We now describe how to perform perturbative computations in the NLSM with $O(N)$ global symmetry in $d= 2+\epsilon$. See e.g. Sec.~4.1.3 of \cite{Henriksson:2022rnm} for a review. 

To solve the constraint in Eq.~\eqref{eq:constraint} explicitly, we identify a component $n^N=\sigma$ along the direction of symmetry breaking and $N-1$ transverse components $n^i=\sqrt{t}\ \pi^i$, with $i=1\dots N-1$, corresponding to the Goldstone bosons. 
Replacing $\sigma=\sqrt{1- t\ \pi^2}$ in the action in Eq.~\eqref{eq:action}, we get an action with canonically normalized $\pi^i$ fields:
\begin{equation}
S=\int d^dx\left(\frac12\partial_\mu\pi  ^i\partial^\mu \pi  ^i+\frac{t}{2}\frac{(\pi  ^i\partial_\mu \pi  ^i)^2}{1-t\pi  ^i\pi  ^i}\right).
\end{equation}

We now replace bare fields and couplings $\pi^i$, $t$ with renormalized ones $\pi^i_R$, $t_R$, by using $\pi  ^i=\sqrt Z_\pi\pi_R  ^i$ and  $t_R=\mu^{-\epsilon}Z_tt$. In minimal subtraction, we have\footnote{We use a different notation with respect to \cite{Henriksson:2022rnm}, as we do not reabsorb the coupling $t$ in the field $\pi$. The renormalization constants $Z_\pi,Z_t$ can be expressed in terms of $Z, Z_1$ of Ref.~\cite{Henriksson:2022rnm} as $Z_\pi=Z/Z_1$ and $Z_t=Z_1$.}
 \be
 \begin{aligned}
     Z_\pi&=1+\delta_\pi=1+ \frac{t_R}{2\pi\epsilon}+\mathcal{O}(t_R^2)\\
     Z_t&=1+\delta_t=1+(N-2) \frac{t_R}{2\pi\epsilon}+\mathcal{O}(t_R^2)\,.
 \end{aligned}
 \label{eq:count}
 \ee
 From this, we can compute the beta function 
\begin{align}
\beta( t_R)&=\epsilon t_R-\frac{(N-2)}{2\pi} t_R^2+\mathcal{O}(\ t_R^2),
\end{align}
which implies the existence of a non-trivial UV fixed-point in $d>2$ for 
\be t_R^*=\frac{2\pi}{N-2}\epsilon+O(\epsilon^2)\,.\label{eq:FP}
\ee 
This fixed point corresponds to the NLSM $O(N)$ CFT and is the main focus of this work.

    \section{Anomalous dimension of composite operators}
    \label{app:andim}    
     Composite operators are obtained by taking products of elementary fields (or their derivatives) at the same space-time point. Even in a renormalized theory, correlation functions involving composite operators have UV divergences, which have to be removed by adding proper counterterms. Similarly to what is done for elementary fields, we can rewrite bare operators in terms of renormalized ones $O=Z_OO_R$ and compute anomalous dimensions as
    \begin{equation}
    \gamma_O=\frac{d \log Z_O}{d \log \mu}\,.
    \label{eq:gamma}
    \end{equation}
    
    Let us consider a bare composite operator containing $m$ elementary fields, so schematically in the form
    \begin{equation}
        O= \partial_{\bar{\mu}^{(1)}}n^{a_1}\dots\partial_{\bar{\mu}^{(m)}}n^{a_m}(x)\,.
        \label{eq:typicalO}
    \end{equation}
    Some or all Lorentz and global indices may be contracted. We assume for simplicity that $O$ is the only operator with given global and Lorentz quantum numbers at this scaling dimension, and therefore that it does not perturbatively mix with other operators under renormalization (if not, we have to generalize this discussion promoting $Z_O$ to a mixing matrix).
    The corresponding renormalization factor is $Z_O=1 + \delta_O$, where the counterterm $\delta_O$ is computed from the divergences of the correlation function of the operator with $m$ elementary fields:
    \begin{equation}
\langle n(x_1)\dots n(x_m) O(x)\rangle,    
\label{eq:corr_O}
    \end{equation}
    where flavor indices of $n$ fields are generic and left implicit. At one loop order, we have
    \begin{equation}\delta_O=\frac{\left.\langle n(x_1)\dots n(x_m) O(x)\rangle_\mathrm{NL} \right|_\mathrm{div}}{\langle n(x_1)\dots n(x_m) O(x)\rangle_\mathrm{L}}\,,
    	\label{eq:delta_O1PI}
    \end{equation}
    where we have denoted with L and NL the leading and the next-to-leading order contributions in $t$ of the correlation function, i.e. the tree level and the one loop terms respectively. As we have assumed that $O$ is the unique operator with this set of quantum numbers and at this scaling dimension, the operator does not mix perturbatively with other operators, and these equations lead to a well-defined anomalous dimension for $O$.

In practice, as explained in 
App.~\ref{app:review}, loop computations in the NLSM are performed by expressing operators in terms of $\pi$ fields. So, it is convenient for our purposes to rewrite the above equations in these terms. 

The simplest is to choose the $n(x_1)\dots n(x_m)$ fields in Eq.~\eqref{eq:corr_O} to be along the $\pi$ direction, $n^a=\sqrt{t} \pi^a$. 
In the transformation from $n$ to $\pi$ some nonlinear factors such as $\sqrt{1-t\pi^2}$ appear. If we expand such factors in Eq.~\eqref{eq:typicalO}, we get an infinite linear combination of operators $O_k$
 \begin{equation}
     O(x) = t^{\#} \tilde O(x),\qquad \tilde O(x)=\sum_{k=0}^\infty O_k(x) t^k\,,
     \label{eq:Otypical_pi}
 \end{equation}
 where $O_k$ contains $m+2k$ fields $\pi$, and we have defined the rescaled operator $\tilde O$ factoring out an overall power of the coupling.
 
We are reduced to studying the rescaled correlation function
   \begin{equation}
\langle \pi(x_1)\dots \pi(x_m) \tilde O(x)\rangle\,,  
\label{eq:corr_O_pi}
\end{equation}
where again flavor indices are left implicit. Eq.~\eqref{eq:delta_O1PI} becomes
\begin{equation}
	\delta_O=\frac{\left.\langle \pi (x_1)\dots \pi(x_m) \tilde O(x)\rangle_\mathrm{NL} \right|_\mathrm{div}}{\langle \pi(x_1)\dots \pi(x_m) \tilde O(x)\rangle_\mathrm{L}}\,.
	\label{eq:delta_O1PIresc}
\end{equation}
$\delta_O$ is composed of two different terms. The first is simply made of the wave-function renormalizations for the elementary fields that compose $O$ and it cancels the divergence of one particle reducible (1PR) diagrams. The second, instead, is needed to make one-particle irreducible (1PI) contributions finite. So we have:
\begin{equation}
\delta_O = \delta_O^{\text{1PI}} + \frac{m}{2} \delta_\pi\,,
\label{eq:ren_comp_pi}
\end{equation}
where $\delta_\pi$ is the wave-function renormalization of $\pi$ defined in Eq.~\eqref{eq:count} and \begin{equation}\delta_O^{1\text{PI}}=\frac{\left.\langle \pi(x_1)\dots \pi(x_m) \tilde O(x)\rangle_\mathrm{NL,1PI} \right|_\mathrm{div}}{\langle \pi(x_1)\dots \pi(x_m) \tilde O(x)\rangle_\mathrm{L}}\,.
\label{eq:delta_O1PI_pi}
  \end{equation}
  
To compute $\delta_O^\mathrm{1P1}$, we express the operator $\tilde O$ in terms of $\pi$ fields, as expressed in Eq.~\eqref{eq:Otypical_pi}.
However, as the terms of the expansion have increasing order in $t$ only the first two terms contribute at one loop, giving 
\begin{equation}
\delta_O^{1\text{PI}}=\frac{\left.(\langle\pi(x_1)\dots\pi(x_m) O_0(x)\rangle_1+t\langle\pi(x_1)\dots\pi(x_m) O_1(x)\rangle_0) \right|_\mathrm{div}}{\langle\pi(x_1)\dots\pi(x_n) O_0(x)\rangle_0}\,,
\label{eq:delta_O1PIbis}
  \end{equation}
where we have denoted with $\langle\dots\rangle_0$ correlation functions without interactions and with $\langle\dots\rangle_1$  1PI correlation functions where the leading interaction vertex
\begin{equation}
   \frac{t}{2} \int d x (\pi^i\partial_\mu \pi^i)^2
\end{equation}
is inserted. At first sight, it is not obvious that the numerator in Eq.~\eqref{eq:delta_O1PIbis} does not introduce mixing terms which would spoil the simplification with the denominator. However, given that this is a rewriting of Eq.~\eqref{eq:delta_O1PI} and that $O$ is a unique operator with a given representation under the full symmetry of the action, we know that this simplification must occur. That it occurs in actual computations serves as a useful check.

Note that, when listing the operators in App.~\ref{app:list_operators}, we have excluded redundant operators, i.e.~proportional to the equations of motion, as they give rise to ultralocal terms when inserted into correlation functions. Specifically, this means that if $O$ satisfies this property, the correlation function in Eq.~\eqref{eq:corr_O_pi} will only contain terms proportional to one of the delta-functions $\delta(x - x_i)$, with $ i = 1, \dots, m $, or a derivative of such a delta-function. Such ultralocal terms would not be visible when comparing say field theory computation with a lattice model which is only defined at distances larger than one lattice spacing, unlike power laws in $x - x_i$ which can be meaningfully compared between field theory and a critical point of a lattice model. In momentum space, the ultralocal terms appear as terms analytic in one of the momenta, and it is easy to detect and drop them.\footnote{We verified this logic by comparing the correlation function $ \langle \pi^a(p_1)\pi^b(p_2)\Box B_{\mu\nu}(q)\rangle $, with $B_{\mu\nu}$ defined in Eq.~\eqref{eq:B_N3}, to the same correlation function where the equations of motion are substituted. As expected, the two differ only by analytic terms in $p_1$ or in $p_2$.} Mixing and renormalization factors of local operators have to be computed once the ultralocal terms have been dropped. For example, if there is no mixing, proportionality in \eqref{eq:delta_O1PI_pi} should hold, but a priori only when the ultralocal terms have been dropped. It turns out that for the computations reported below this subtlety does not play a role, but it has to be kept in mind in more general situations.

\subsection{Check of $B_{\mu\nu}$ scaling dimension}
\label{sec:andimB}
In Sec.~\ref{sec:presentation} we showed that the operator $B$ defined in Eq.~\eqref{eq:B_N>3} has a protected dimension equal to $N-1$. 
As a check, we perform a loop computation to verify this statement for $N=3$ at the order-$\epsilon$. 
We start by expanding the operator in terms of the Goldstone bosons $\pi$ and thus obtain\footnote{This operator differs by a factor $t$ with respect to the operator in Eq.~\eqref{eq:B_N>3}, but as discussed this rescaling does not affect the final result for the dimension of the renormalized operator. 
}
\begin{equation}
    B_{\mu\nu}=(B_0)_{\mu\nu}+t(B_1)_{\mu\nu}+O(t^2) \,,
    \label{eq:B_N3}
\end{equation}
with
\begin{equation}
    \begin{aligned}(B_0)_{\mu\nu}&=2\partial_{[\mu}\pi^1\partial_{\nu]}\pi^2\,,\\(B_1)_{\mu\nu}&=\partial_{[\mu}\pi^1\partial_{\nu]}\pi^2(\pi^i \pi^i)\,.
    \end{aligned}
\end{equation}
At leading order, as $\pi$ and $t$ have classic dimension equal to $\epsilon/2$ and $-\epsilon$ respectively, we have
\be
\Delta_B= 2+\epsilon+\gamma_B+\mathcal{O}(\epsilon^2)\,,
\label{eq:delta_B}
\ee
where $\gamma_B$ can be computed following the procedure explained in the previous section. 

We then  consider the correlation function 
\begin{equation}
G^{ij}_{\mu\nu}(x_1,x_2,x)=\langle\pi^i(x_1)\pi^j(x_2) {B}_{\mu\nu}(x)\rangle
\end{equation}
and decompose it into leading and subleading contributions 
\begin{equation}
\begin{aligned}
&( G^{ij}_{\mu\nu})_\mathrm{L}=\langle\pi^i(x_1)\pi^j(x_2) {(B_0)}_{\mu\nu}(x)\rangle_\mathrm{0}\\
&( G^{ij}_{\mu\nu})_\mathrm{NL,1PI}=\langle\pi^i(x_1)\pi^j(x_2) {(B_0)}_{\mu\nu}(x)\rangle_1+t\langle\pi^i(x_1)\pi^j(x_2) {(B_1)}_{\mu\nu}(x)\rangle_\mathrm{0}\,,
\end{aligned}
\label{eq:correlationB}
\end{equation}
which correspond to the denominator and the numerator of Eq.~\eqref{eq:delta_O1PIbis}.
 The leading order is easily computed by taking Wick contractions. In momentum space, we find
\begin{equation}
  ( G^{ij}_{\mu\nu})_\mathrm{L}=\langle\pi^i(p_1)\pi^j(p_2) (B_0)_{\mu\nu}(q)\rangle_\mathrm{L}
    ={ 2\epsilon^{ij} }\delta(q-p_1-p_2)(p_{1\mu}p_{2\nu}-p_{2\mu}p_{1\nu}) G(p_1)G(p_2)\,,
    \label{eq:correlationBL}
\end{equation}
where $G(p)$ is the propagator for the $\pi$ field in momentum space and simply reads
\begin{equation}
   G(p)=\frac{1}{p^2+H} \,,
\end{equation}
with $H$ being an IR regulator which we eventually take to zero.

At the next-to-leading order, the two contributions in Eq.~\eqref{eq:correlationB} correspond to the 1PI diagrams (a) and (b) presented in Fig.~\ref{fig:diagrams}. Summed together, they give
\begin{equation}
    ( G^{ij}_{\mu\nu})_\mathrm{NL,1PI}= 2t(G^{ij}_{\mu\nu})_\mathrm{L}\int d^d k \frac{1}{k^2+H}\ \,,
\end{equation}
where $(G^{ij}_{\mu\nu})_\mathrm{L}$ is defined in Eq.~\eqref{eq:correlationBL}. Doing the integral with dimensional regularization, replacing the bare coupling with the renormalized one,  and focusing on the divergent term, we get
\begin{equation}
     \left.( G^{ij}_{\mu\nu})_\mathrm{NL,1PI}\right|_\mathrm{div}=-\frac{t_R}{\pi \epsilon}(G^{ij}_{\mu\nu})_\mathrm{L}\,,
\end{equation}
which implies
\begin{equation}
\delta_{{B}}^{ 1\text{PI}}=-\frac{t_R}{\pi\epsilon}\,.
\end{equation}
Replacing in Eq.~\eqref{eq:ren_comp_pi}, we get  \begin{equation}
Z_B=1+\delta_B\,, \quad \text{with } \quad \delta_B=\delta_B^{1\text{PI}}+\delta_\pi\,,
\end{equation}
 which gives
\begin{equation}
    \gamma_B(t_R)=-\frac{1}{2\pi} t_R\,,
\end{equation}
and therefore  $\gamma_B(t_R^*)=-\epsilon$ at the fixed point corresponding to the NLSM $O(N)$ CFT. Substituting into \eqref{eq:delta_B}, we verify that the operator $B_{\mu\nu}$ has a protected dimension equal to 2.
\begin{figure}
    \centering

    \begin{minipage}{0.3\textwidth}
        \centering
\begin{tikzpicture}
    \node[circle, inner sep=1.5pt] (D) at (-2, 0) {};
    \node[circle, fill=black, inner sep=1.5pt] (A) at (-1, 0) {};
    \node[circle, fill=black, inner sep=1.5pt] (B) at (1, 0) {};

    \draw[thick] (A) to[out=60, in=120] (B);
    \draw[thick] (A) to[out=-60, in=-120] (B);

    \draw[thick] (B) -- (2, 0.7);
    \draw[dashed] (D) -- (A);
    \draw[thick] (B) -- (2, -0.7);
\end{tikzpicture}
\caption*{(a)}\label{fig:2a}\end{minipage}\hspace{1cm}\begin{minipage} {0.3\textwidth}
\centering
\begin{tikzpicture}
    \node[circle, inner sep=1.5pt] (D) at (-2, 0) {};
    \node[circle, fill=black, inner sep=1.5pt] (A) at (-1, 0) {};


    \draw[thick] (A) -- (1, 0.7);
    \draw[dashed] (D) -- (A);
    \draw[thick] (A) -- (1, -0.7);
     \draw[thick] (A) .. controls (-0, 1) and (-2, 1) .. (A);
\end{tikzpicture} 
\caption*{(b)}
    \end{minipage}
    \caption{1PI one-loop diagrams contributing to the correlation function $G_{\mu\nu}^{ij}$. Here the dashed outgoing line represents the momentum $q$ of the composite operator, and the solid outgoing lines correspond to momenta $p_1$ and $p_2$ of the external pion fields.}
    \label{fig:diagrams}
\end{figure}

\bibliographystyle{utphys}
\bibliography{NLSM}

\end{document}